%% file: paper.tex
\documentclass[letterpaper,twocolumn,10pt]{article}

\input{style/boilerplate}
\usepackage{style/usenix}
\usepackage{listings}

\def\restler{{\em REST-ler}\xspace}
\def\gitlab{GitLab\xspace}
\def\nGitlabAPIGroups{63\xspace}
\def\nGitlabSwaggerRequests{358\xspace}
\def\nBootCodeCoverage{14,468\xspace}
\def\nUnicornWorkers{10\xspace}
\def\nPostGREWorkers{10\xspace}
\def\nUnicornMem{900MB\xspace}

\newcommand{\HiddenNote}[1]{}

\begin{document}

\title{REST-ler: Automatic Intelligent REST API Fuzzing}


\author{Vaggelis Atlidakis\thanks{The work of this author was mostly
    done while visiting Microsoft Research.}\\Columbia University \and Patrice Godefroid\\Microsoft Research \and Marina Polishchuk\\Microsoft Research}

\maketitle

\input{sections/00-abstract}

\input{sections/01-introduction}
\input{sections/02-overview}
\input{sections/03-restler}
\input{sections/04-0-evaluation}

\input{sections/05-case-studies}
\input{sections/06-related-work}

\input{sections/07-conclusion}

\renewcommand{\baselinestretch}{0.9}
{
  \small
  \bibliographystyle{abbrv}
  \bibliography{paper}
}

\end{document}

%% file: style/boilerplate.tex
\usepackage{times}
\usepackage{graphicx}
\usepackage{subfigure}
\usepackage{array} 
\usepackage{amsmath,amsfonts,amssymb,amsthm}
\usepackage{color}
\usepackage[hyphens]{url}
\usepackage{listings} 
\usepackage{xspace} 
\usepackage{booktabs} 
\usepackage[font=small]{caption}
\usepackage{courier} 
\usepackage{verbatim}
\usepackage[usenames,dvipsnames,svgnames,table]{xcolor}
\usepackage{calc}
\usepackage{amsthm}
\usepackage{algorithm, algpseudocode}
\usepackage{multirow}

\let\oldenumerate\enumerate
\renewcommand{\enumerate}{
  \oldenumerate
  \setlength{\itemsep}{.0pt}
  \setlength{\parskip}{0pt}
  \setlength{\parsep}{0pt}
}

\let\olditemize\itemize
\renewcommand{\itemize}{
  \olditemize
  \setlength{\itemsep}{1pt}
  \setlength{\parskip}{0pt}
  \setlength{\parsep}{0pt}
}

\newtheorem*{proposition*}{Proposition}
\theoremstyle{definition}

\makeatletter
\renewcommand{\ALG@beginalgorithmic}{\footnotesize}
\makeatother

\newcommand{\ie}{{\em i.e.,~}}
\newcommand{\eg}{{\em e.g.,~}}

\def\F{Figure~}
\def\T{Table~}

\newcommand{\ar}[3]{} 
\include{latexdefs}
\include{mathnotation}

\newcommand{\heading}[1]{{\vspace{0pt}\noindent\bf{#1}}} 

{\makeatletter
 \gdef\xxxmark{%
   \expandafter\ifx\csname @mpargs\endcsname\relax 
     \expandafter\ifx\csname @captype\endcsname\relax 
       \marginpar{\textcolor{red}{xxx~}}
     \else
       \textcolor{red}{xxx~}
     \fi
   \else
     \textcolor{red}{xxx~}
   \fi}
 \gdef\xxx{\@ifnextchar[\xxx@lab\xxx@nolab}
 \long\gdef\xxx@lab[#1]#2{{\bf [\xxxmark \textcolor{red}{#2} ---{\sc #1}]}}
 \long\gdef\xxx@nolab#1{{\bf [\xxxmark \textcolor{red}{#1}]}}
}

{\makeatletter
 \gdef\edit{\@ifnextchar[\edit@lab\edit@nolab}
 \long\gdef\edit@lab[#1]#2{[\textcolor{red}{#2} ---{\sc #1}]}
 \long\gdef\edit@nolab#1{[\textcolor{red}{#1}]}
}

\newcommand{\ignore}[1]{}

\definecolor{grey}{rgb}{0.5,0.5,0.5}

\lstdefinelanguage{yaml}{
  alsoletter=/,
  ndkeywords={id, post, body, string, /blog/posts/, /api, string,
              integer},
  ndkeywordstyle=\color{purple},
  basicstyle=\footnotesize,
}

\lstdefinelanguage{restler}{
  keywords={requests, Request, set_var, writer, dependencies, parser,
            post_send, restler_static, restler_fuzzable},
  keywordstyle=\color{darkgray}\bfseries,
  stringstyle=\color{purple},
  morestring=[b]",
  basicstyle=\footnotesize,
}

\lstdefinelanguage{networklog}{
  alsoletter=/0123456789.,
  ndkeywords={POST, /api/blog/posts/, HTTP/1.1, body, id,
              sampleString, 5889}
  ndkeywordstyle=\color{purple},
  basicstyle=\footnotesize,
}

\lstdefinelanguage{bucketlog}{
  alsoletter=/0123456789.,
  ndkeywords={HTTP/1.1, string, /api/v4/projects/projectid/repository/commits,
              /api/v4/projects, POST},
  ndkeywordstyle=\color{purple},
  keywords={requests, Request, set_var, writer, dependencies, parser,
            post_send, restler_static, restler_fuzzable},
  keywordstyle=\color{darkgray}\bfseries,
  basicstyle=\footnotesize,
}

\newcommand{\code}[1]{\tt{#1}}

%% file: sections/00-abstract.tex
\begin{abstract}
Cloud services have recently exploded with the advent of powerful
cloud-computing platforms such as Amazon Web Services and Microsoft
Azure. Today, most cloud services are accessed through REST APIs, and
Swagger is arguably the most popular interface-description language
for REST APIs. A Swagger specification describes how to access a cloud
service through its REST API (\eg what requests the service can
handle and what responses may be expected).

This paper introduces \restler, the first automatic intelligent REST
API security-testing tool. \restler analyzes a Swagger specification
and generates tests that exercise the corresponding cloud
service through its REST API. Each test is defined as a sequence of
requests and responses. \restler generates tests intelligently by (1)
inferring {\em dependencies among request types} declared in the
Swagger specification (\eg inferring that ``a request B should not
be executed before a request A'' because B takes as an input argument
a resource-id $x$ returned by A) and by (2) analyzing {\em dynamic
feedback} from responses observed during prior test executions in
order to generate new tests (\eg learning that ``a request C after a
request sequence A;B is refused by the service'' and therefore
avoiding this combination in the future). We show that these two
techniques are necessary to thoroughly exercise a service under test
while pruning the large search space of possible request sequences. We
also discuss the application of
\restler to test GitLab, a large popular open-source self-hosted Git
service, and the new bugs that were found.

\end{abstract}

%% file: sections/01-introduction.tex
\section{Introduction}
\label{sec:introduction}

Over the last decade, we have seen an explosion in cloud services
for hosting software applications (Software-as-a-Service), for data
processing (Platform-as-a-Service), and for providing general
computing infrastructure (Infrastructure-as-a-Service). Today, most
cloud services, such as those provided by Amazon Web Services and
Microsoft Azure, are programmatically accessed through REST
APIs~\cite{REST} by third-party applications~\cite{REST-book} and
other services~\cite{services-book}. Meanwhile, Swagger~\cite{Swagger}
(recently renamed OpenAPI) has arguably become the most popular
interface-description language for REST APIs. A Swagger specification
describes how to access a cloud service through its REST API,
including what requests the service can handle and what responses may
be expected in what format.

Tools for automatically testing cloud services via their REST APIs and
checking whether those services are reliable and secure are still in
their infancy. The most sophisticated testing tools currently
available for REST APIs capture live API traffic, and then parse, fuzz
and replay the traffic with the hope of finding
bugs~\cite{appspider,qualysWAS,boofuzz,tnt-fuzzer,apifuzzer}. Many of
these tools were born as extensions of more established web-site
testing and scanning tools (see Section~\ref{sec:relwork}). Since
these REST API testing tools are all recent and not widely used, it is
currently unknown how effective they are in finding bugs and how
security-critical those bugs are.

In this paper, we introduce \restler, the first automatic intelligent
REST API fuzzing tool. Fuzzing~\cite{fuzzing-book} means automatic
test generation and execution with the goal of finding security
vulnerabilities. Unlike other REST API testing tools, \restler
performs a lightweight static analysis of an entire Swagger
specification, and then generates and executes tests that exercise the
corresponding cloud service through its REST API. Each test is defined
as a sequence of requests and responses. \restler generates tests {\em
intelligently} by
\begin{enumerate}
\topsep0pt
\itemsep0pt
\item inferring {\em dependencies among request types}
declared in the Swagger specification (\eg inferring that a resource
included in the response of a request A is necessary as input argument
of another request B, and therefore that A should be executed before
B), and by
\item analyzing {\em dynamic feedback} from responses observed
during prior test executions in order to generate new tests (\eg
learning that ``a request C after a request sequence A;B is refused by
the service'' and therefore avoiding this combination in the future).
\end{enumerate}
We present empirical evidence that these two techniques (partly
inspired by prior work on API testing for object-oriented
programs~\cite{randoop}) are necessary to thoroughly exercise a
service under test while pruning the large search space defined by all
possible request sequences. \restler also implements several search
strategies (some inspired by prior work on model-based
testing~\cite{YL91}), and we compare their effectiveness while fuzzing
GitLab~\cite{gitlab}, a large popular open-source self-hosted Git
service with a complex REST API. During the course of our experiments,
we found several new bugs in GitLab, including security-relevant ones
(see Section~\ref{sec:case-studies}).

In summary, this paper makes the following contributions:
\begin{itemize}
\item We introduce \restler, the first automatic intelligent fuzzing tool for REST APIs
which analyzes a Swagger specification, automatically infers
dependencies among request types, generates tests defined as request
sequences satisfying those dependencies, and dynamically learns what
request sequences are valid or invalid by 
analyzing the service responses to those tests.
\item We present detailed experimental evidence showing that the techniques used in
\restler are necessary for effective automated REST API fuzzing.
\item We also present experimental results obtained with three different strategies
for searching the large search space defined by all possible request
sequences, and discuss their strengths and weaknesses.
\item We present a detailed case study with GitLab, a large popular open-source self-hosted
Git service, and discuss several new bugs found so far and their severity.
\end{itemize}

This paper is organized as follows. In the next Section, we describe
Swagger specifications and how they are processed by \restler. In
Section~\ref{sec:restler}, we present the main test-generation
algorithm used in \restler, and discuss different search strategies
and other implementation details. In Section~\ref{sec:evaluation}, we
present experimental results evaluating the effectiveness of the
test-generation techniques and search strategies implemented in
\restler. In Section~\ref{sec:case-studies}, we discuss several new
bugs found in GitLab during the course of this work. We discuss
related work in Section~\ref{sec:relwork} and conclude
in Section~\ref{sec:conclusion}.

%% file: sections/02-overview.tex
\section{Processing API Specifications}
\label{sec:overview}

In this paper, we consider cloud services accessible through REST APIs
described with a Swagger specification. A Swagger specification
describes how to access a cloud service through its REST API (\eg
what requests the service can handle and what responses may be
expected). A client program can send messages, called {\em requests},
to a service and receive messages back, called {\em responses}. Such
messages are sent over the HTTP protocol. Given a Swagger
specification, open-source Swagger tools can automatically generate a
web UI that allows users to view the documentation and interact
with the API via a web browser.

\begin{figure}[t]
    \centering
    \includegraphics[width=0.48\textwidth]{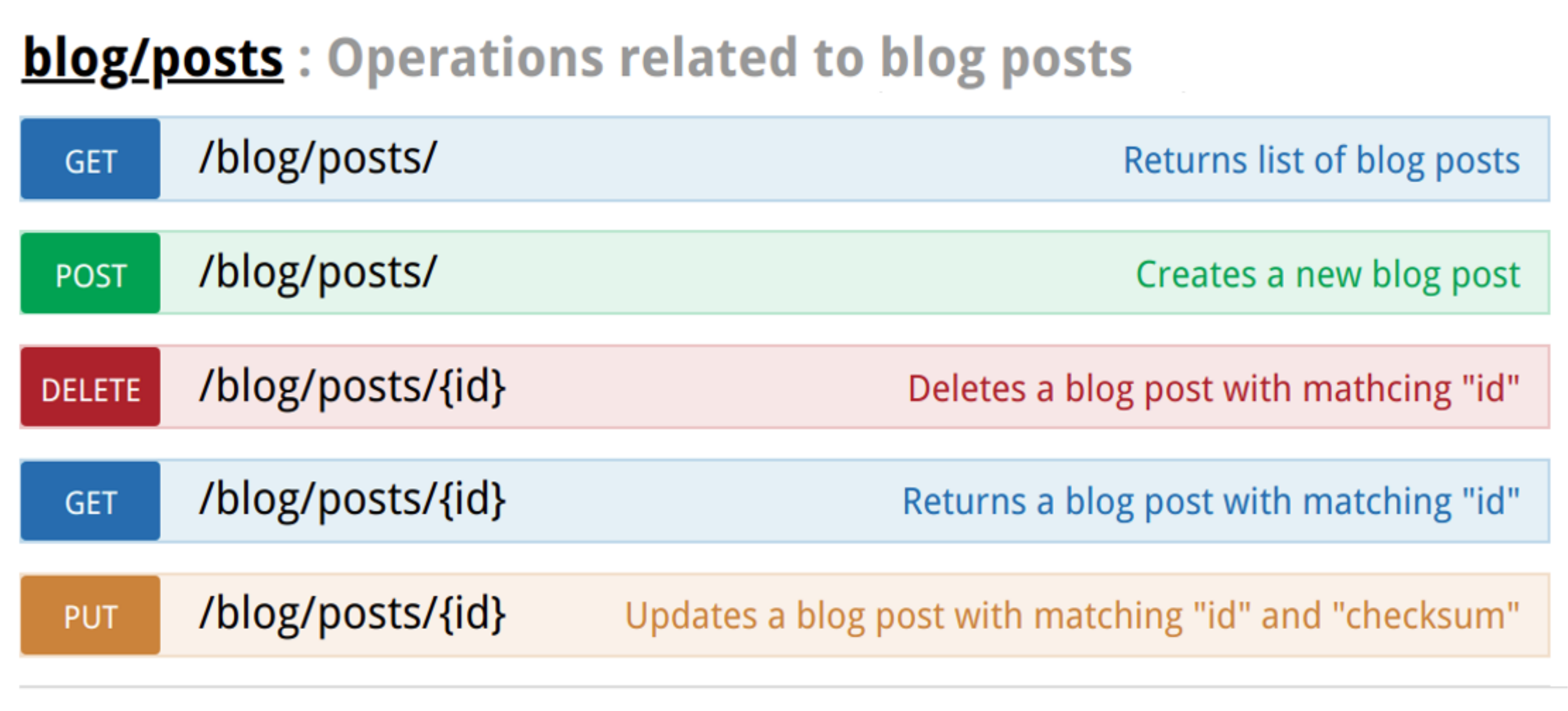}
    \caption{
        {\bf Swagger Specification of Blog Posts Service}
    }
    \label{fig:demoserver_api}
\end{figure}

An example of Swagger specification in web-UI form is shown in
Figure~\ref{fig:demoserver_api}. This specification describes five
types of requests supported by a simple service for hosting blog
posts. This service allows users to create, access, update and
delete blog posts. In a web browser, clicking on any of these five
request types expands the description of the request type.

\newbox\swaggerSpec
\begin{lrbox}{\swaggerSpec}
\begin{lstlisting}[linewidth=110pt,mathescape=true, showstringspaces=false, basicstyle=\footnotesize, language=yaml, commentstyle=\color{olive}, keywordstyle=\color{blue}, frame=none]
basePath: '/api'
swagger: '2.0'
definitions:
 "Blog Post":
  properties:
   body:
    type: string
   id:
    type: integer
  required:
  -body
  type: object

paths:
 "/blog/posts/"
  post:
   parameters:
   -in: body
    name: payload
    required: true
   schema:
    ref: "/definitions/Blog Post"
\end{lstlisting}
\end{lrbox}

\newbox\restlerSpec
\begin{lrbox}{\restlerSpec}
\begin{lstlisting}[linewidth=110pt,language=restler,frame=none]
from restler import requests
from restler import dependencies

def parse_posts(data):
 post_id = data["id"]
 dependencies.set_var(post_id)

request = requests.Request(
 restler_static("POST"),
 restler_static("/api/blog/posts/"),
 restler_static("HTTP/1.1"),
 restler_static("{"),
 restler_static("body:"),
 restler_fuzzable("string"),
 restler_static("}"),
 'post_send': {
   'parser': parse_posts,
   'dependencies': [
      post_id.writer(),
   ]
 }
)
\end{lstlisting}
\end{lrbox}

\begin{figure}[t]
    \begin{minipage}{0.25\textwidth}
        \usebox\swaggerSpec
    \end{minipage}
    \begin{minipage}{0.22\textwidth}
        \usebox\restlerSpec
    \end{minipage}
    \caption{
        {\bf Swagger Specification and Automatically Derived \restler Grammar.}
        Shows a snippet of Swagger specification in YAML format
        (left) and the corresponding grammar generated by
        \restler (right).
    }
    \label{fig:demoserver_swagger_restler_snippets}
\end{figure}


For instance, selecting the second request, which is a POST request,
reveals text similar to the left of
Figure~\ref{fig:demoserver_swagger_restler_snippets}. This text is in
YAML format and describes the exact syntax expected for that specific
request and its response. In this case, the {\tt definition} part of
the specification indicates that an object named {\tt body} of type {\tt string} is
required and that an object named {\tt id} of type {\tt integer} is optional (since
it is not required). The {\tt path} part of the specification
describes the HTTP-syntax for a POST request of this type as well as
the format of the expected response.

From such a specification, \restler automatically constructs the
test-generation grammar shown on the right of
Figure~\ref{fig:demoserver_swagger_restler_snippets}. This grammar is
encoded in executable {\tt python} code. It mainly consists of code to
generate an HTTP request, of type POST in this case. Each command {\tt
restler\_static} simply appends the string it takes as argument
without modifying it. In contrast, the command {\tt restler\_fuzzable}
takes as argument a value type (like {\tt string} in this example) and
replaces it by one value of that type taken from a (small) {\em
dictionary} of values for that type. How dictionaries are defined and
how values are selected is discussed in the next section.

\newbox\restlerHTTPSample
\begin{lrbox}{\restlerHTTPSample}
\begin{lstlisting}[linewidth=160pt,language=networklog]
Sending: POST /api/blog/posts/ HTTP/1.1
Accept: application/json
Content-Type: application/json
Host: localhost:8888
{"body":"sampleString"}

Received: HTTP/1.1 201 CREATED
Content-Type: application/json
Content-Length: 37
Server: Werkzeug/0.14.1 Python/2.7.12
Date: Sun, 01 Apr 2018 05:10:32 GMT
{"body": "sampleString", "id": 5889}
\end{lstlisting}
\end{lrbox}

\begin{figure}[t]
    \center
        \usebox\restlerHTTPSample
	\vspace{-5pt}
    \caption{
        {\bf \restler Trace of HTTP Request and Response.}
        Shows a network-layer \restler trace with a POST
        request creating a blog post and the respective response.
    }
    \label{fig:demoserver_http_samples}
\end{figure}

The grammar on the right also includes code to process the expected
response of the request. In this case, the response is expected to
return a new object named {\tt id} of type {\tt integer}.
Using the {\tt schema} specified on the left, \restler
automatically generates the function {\tt parse\_posts} shown on the
right.
\F\ref{fig:demoserver_http_samples} shows an example of a HTTP-level trace of a single POST request
generated by \restler for the blog posts service and the corresponding response.

By similarly analyzing the other request types described in this
Swagger specification, \restler will infer automatically that {\tt id}s
returned by such POST requests are necessary to generate well-formed
requests of the last three request types shown in
\F\ref{fig:demoserver_api} which each requires an {\tt
{id}}. These {\em dependencies} are extracted by \restler when
processing the Swagger specification and are later used for test
generation, as described next.

%% file: sections/03-restler.tex
\section{\restler}
\label{sec:restler}

\subsection{Test Generation Algorithm}

\begin{figure}
\begin{lstlisting}[frame=none, mathescape=true, showstringspaces=false, basicstyle=\footnotesize, language=Python, numbers=left, numbersep=5pt, numberstyle=\tiny, commentstyle=\color{purple}, keywordstyle=\color{blue}]
Inputs: swagger_spec, maxLength
# Set of requests parsed from the Swagger API spec
reqSet = PROCESS(swagger_spec)
# Set of request sequences (initially empty)
seqSet = {}
# Main loop: iterate up to a given maximum sequence length
n = 1
while (n =< maxLength):
  seqSet = EXTEND(seqSet, reqSet)
  seqSet = RENDER(seqSet)
  n = n + 1
# Extend all sequences in seqSet by appending
# new requests whose dependencies are satisfied
def EXTEND(seqSet, reqSet):
  newSeqSet = {}
  for seq in seqSet:
    for req in reqSet:
      if DEPENDENCIES(seq, req):
        newSeqSet = newSeqSet + concat(seq, req)
  return newSeqSet
# Concretize all newly appended requests using dictionary values,
# execute each new request sequence and keep the valid ones
def RENDER(seqSet):
  newSeqSet = {}
  for seq in seqSet:
    req = last_request_in(seq)
    $\vec{V}$ = tuple_of_fuzzable_types_in(req)
    for $\vec{v}$ in $\vec{V}$:
      newReq = concretize(req, $\vec{v}$)
      newSeq = concat(seq, newReq)
      response = EXECUTE(newSeq)
      if response has a valid code:
        newSeqSet = newSeqSet + newSeq
      else:
        log error
  return newSeqSet
# Check that all objects referenced in a request are produced
# by some response in a prior request sequence
def DEPENDENCIES(seq, req):
  if CONSUMES(req) $\subseteq$ PRODUCES(seq):
    return True
  else:
    return False
# Objects required in a request
def CONSUMES(req):
  return object_types_required_in(req)
# Objects produced in the responses of a sequence of requests
def PRODUCES(seq):
  dynamicObjects = {}
  for req in seq:
    newObjs = objects_produced_in_response_of(req)
    dynamicObjects = dynamicObjects + newObjs
  return dynamicObjects
\end{lstlisting}
\caption{{\bf Main Algorithm used in \restler.}}
\label{fig:main-algorithm}
\end{figure}

The main algorithm for test generation used by \restler is shown in
Figure~\ref{fig:main-algorithm} in {\tt python}-like notation. It
starts (line~3) by processing a Swagger specification as discussed in the
previous section. The result of this processing is a set of request
types, denoted {\tt reqSet} in Figure~\ref{fig:main-algorithm}, and of
their dependencies (more on this later).

The algorithm computes a set of request sequences, denoted {\tt
seqSet} and initially empty (line~5). At each iteration of its main
loop (line~8), the algorithm computes {\em all valid} request
sequences {\tt seqSet} of length $n$, starting with $n=1$ before
moving to $n+1$ and so on until a user-specified {\tt maxLength} is
reached. Computing {\tt seqSet} is done in two steps.

First, the set of valid request sequences of length $n-1$ is {\em
extended} (line~8) by appending at the end of each sequence one new
request whose dependencies are satisfied, for all possible requests,
as described in the EXTEND function (line~14). The function
DEPENDENCIES (line~39) checks that all the object types required in
the last request, denoted by {\tt CONSUMES(req)}, are produced by some
response in the request sequence preceding it, denoted by {\tt
PRODUCES(seq)}. If all the dependencies are satisfied, the new
sequence of length $n$ is retained (line~19), otherwise it is
discarded.

Second, each newly-extended request sequence whose dependencies are
satisfied is {\em rendered} (line~10) one by one as described in the
RENDER function (line~23). For every newly-appended request (line~26),
the list of all fuzzable primitive types in the request is computed
(line~27) (those are identified by {\tt restler\_fuzzable} in the code
shown on the right of
Figure~\ref{fig:demoserver_swagger_restler_snippets}). Then, each
fuzzable primitive type in the request is replaced by one concrete
value of that type taken out of a finite (and small) dictionary of
values. The function RENDER generates all possible such combinations
(line~28). Each combination thus defines a fully-defined request {\tt
newReq} (line~29) which is HTTP-syntactically correct. The function
RENDER then {\em executes} this new request sequence (line~31), and
checks its response: if the response has a valid return code (defined
here as any code in the {\tt 200} range), the new request sequence is
``valid'' and retained (line~33), otherwise it is discarded and the
received error code is logged for further analysis and debugging.

More precisely, the function EXECUTE executes each request in a
sequence request one by one, each time checking that the response is
valid, extracting and memoizing dynamic objects (if any), and
providing those in subsequent requests in the sequence if needed, as
determined by the dependency analysis; the response returned by
function EXECUTE in line~31 refers to the response received for the
last, newly-appended request in the sequence.  Note that if a request
sequence produces more than one dynamic object of a given type, the
function EXECUTE will memoize all of those objects, but will provide
them later when needed by subsequent requests in the exact order in
which they are produced; in other words, the function EXECUTE will not
try different ordering of such objects. If a dynamic object is passed
as argument to a subsequent request and is ``destroyed'' after that
request, that is, it becomes unusable later on, \restler will detect
this by receiving an invalid response ({\tt 400} or {\tt 500} error
code) when attempting to reuse that unusable object, and will then
discard that request sequence.

For each fuzzable primitive type, the algorithm of
Figure~\ref{fig:main-algorithm} uses a small set of values of that
type, called {\em dictionary}, and picks one of these values in order
to {\em concretize} that fuzzable type (lines~27-29). For instance,
for fuzzable type {\tt integer}, \restler might use a small dictionary
with the values 0, 1, and -10, while for fuzzable type {\tt string}, a
dictionary could be defined with the values ``sampleString'', the
empty string and one very long fixed string. The user defines those
dictionaries.

By default, the function RENDER of Figure~\ref{fig:main-algorithm}
generates {\em all} possible combinations of dictionary values for
every request with several fuzzable types (see line~28). For large
dictionaries, this may result in astronomical numbers of
combinations. In that case, a more scalable option is to randomly
sample each dictionary for one (or a few) values, or to use {\em
combinatorial-testing} algorithms~\cite{combinatorial-testing} for
covering, say, every dictionary value, or every pair of values, but
not every $k$-tuple.  In the experiments reported later, we used small
dictionaries and the default RENDER function shown in
Figure~\ref{fig:main-algorithm}.

The function EXTEND of Figure~\ref{fig:main-algorithm} generates {\em
all} request sequences of length $n+1$ whose dependencies are
satisfied. Since $n$ is incremented at each iteration of the main loop
of line~8, the overall algorithm performs a {\em breadth-first search}
(BFS) in the search space defined by all possible request
sequences. In Section~\ref{sec:evaluation}, we report experiments
performed also with two additional search strategies: BFS-Fast and
RandomWalk.

\heading{\bf BFS-Fast.}  In function EXTEND, the loops of line~17 and line~18
are swapped in such a way that every request {\tt req} in {\tt reqSet}
is appended only once to some request sequence in {\tt seqSet} in line~19,
resulting in a smaller set {\tt newSeqSet} which covers (\ie
includes at least once) every request but does not generate all valid
request sequences. Like BFS, BFS-Fast thus still provides full
grammar-coverage at each iteration of the main loop in line~8, but it
generates fewer request sequences, which allows it to go deeper more
quickly than BFS.

\heading{\bf RandomWalk.}  In function EXTEND, the two loops of line~17 and
line~18 are eliminated; instead, the function now returns a single new
request sequence whose dependencies are satisfied, and generated by
{\em randomly} selecting one request sequence {\tt seq} in {\tt
seqSet} and one request in {\tt reqSet}. (The function randomly
chooses such a pair until all the dependencies of that pair are
satisfied.) This search strategy will therefore explore the search
space of possible request sequences deeper more quickly than BFS or
BFS-Fast. When RandomWalk can no longer extend the current request
sequence, it restarts from scratch from an empty request
sequence. (Since it does not memoize past request sequences between
restarts, it might regenerate the same request sequence again in the
future.)

\subsection{Implementation Details}

We have implemented \restler with 2,230 lines of python code. Its
functionality is split into four modules: the main application entry point,
the parser and compiler module, the core fuzzing engine, and the logging module.

The main application entry-point is responsible for starting
a fuzzing session according to a set of configuration parameters controlling:
the Swagger specification which should be used to derive an
input grammar and perform fuzzing;
the desired search strategy, including BFS, BFS-Fast,
and RandomWalk; the maximum sequence length and the maximum fuzzing
time; the HTTP status codes indicating errors (\eg {\tt 500});
the dictionary of fuzzing mutations to be used when rendering fuzzable primitive types;
and the port and IP-address of the fuzzing target along with any additional
authorization tokens.

The parser and compiler module is responsible for parsing a Swagger
specification and generating a \restler grammar for fuzzing the
target service. In the absence of a Swagger specification, the user can manually
provide a \restler grammar to be used for fuzzing.

The core engine module implements the algorithm of
\F\ref{fig:main-algorithm},
and is responsible for rendering API requests and for composing sequences of
requests using any of the supported search strategies.
The rendered request sequences are sent
to the target service using {\code send} on {\code python} sockets.
Similarly, the corresponding response is received using {\code recv} on {\code
python} sockets.

Finally, the logging module monitors
client/service interactions and traces all
messages exchanged via {\code python} sockets.
Sequences of requests sent to the target service, along with the corresponding
HTTP status codes received from the service, are persistently stored and
inspected for errors and bug detection.

\subsection{Current Limitations}
Currently \restler only supports token-based authorization, such as OAUTH~\cite{oauth},
and there in no support for operations that involve web-UI interactions.
Moreover, \restler does not support requests for API endpoints that
depend on server-side redirects  (\eg {\tt 301} ``Moved Permanently'',
{\tt 303} ``See Other'', and {\tt 307} ``Temporary Redirect''). Finally,  our current
\restler
prototype can only find bugs defined as unexpected
HTTP status codes, namely {\tt 500} ``Internal Server Error''.
Such a simple test oracle cannot detect vulnerabilities that are
not visible though HTTP status codes (\eg ``Information Exposure'' and others).
Despite these limitations, \restler is already useful in finding
security-relevant bugs, as will be discussed in Section~\ref{sec:case-studies}.

%% file: sections/04-0-evaluation.tex
\section{Evaluation}
\label{sec:evaluation}
We present experimental results obtained with \restler
that answer the following questions:
\begin{enumerate}
    \item[\underline{Q1:}] Are both inferring dependencies among
    request types and analyzing dynamic feedback necessary for
    effective automated REST API fuzzing?
        (Section~\ref{sec:eval:benchmarks})
    \item[\underline{Q2:}] Are tests generated by \restler
        exercising deeper service-side logic as sequence length increases?
        (Section~\ref{sec:eval:gitlab})
    \item[\underline{Q3:}] What search strategy should be used in \restler?
        (Section~\ref{sec:eval:fuzzing-schemes})
    \vspace{-4pt}
\end{enumerate}
We answer the first question (Q1) using a simple Model-View-Controller
(MVC) blog posts service with a REST API. We answer (Q2), and (Q3) using
\gitlab, an open-source, production-scale
\footnote{
    GitLab~\cite{gitlab}
    is used by more than 100,000 organizations, has millions of
    users, and has currently a $2/3$ market share of the self-hosted Git
    market~\cite{gitlab-stats}.
}
web service for self-hosted Git. We conclude the evaluation by
discussing in Section~\ref{sec:eval:usability} how to bucketize (\ie
group together) the numerous bugs that can be reported by \restler in
order to facilitate their analysis. Afterwards, we move on to
Section~\ref{sec:case-studies} where we discuss new bugs found in GitLab.

\input{sections/04-1-experimental-setup}
\input{sections/04-2-demo-server}
\input{sections/04-3-gitlab}
\input{sections/04-4-fuzzing-schemes}
\input{sections/04-5-usability}

%% file: sections/04-1-experimental-setup.tex
\subsection{Experimental Setup}
\label{sec:subsec:setup}
\heading{Blog Posts Service.}
We answer (Q1) using a simple blog posts service, written in $189$ lines of
python code using the Flask web framework~\cite{flask} and following
the MVC web development paradigm.
Every blog post is persistently stored in a SQLite~\cite{sqlite} database and
has a user-assigned body, an automatically-assigned post id (primary key),
and an automatically-derived checksum (SHA-1) of the blog post's body.
The service's functionality is exposed over a REST API with a Swagger specification
shown in Figure~\ref{fig:demoserver_api}. This API contains
five request types:
(i) GET on {\code /posts} : returns all blog posts currently
registered;
(ii) POST on {\code /posts} : creates a new blog post (body:
the text of the blog post);
(iii) DELETE {\code /posts/id} :
deletes a blog post;
(iv) GET {\code posts/id} : returns the body and the checksum of an individual
blog post;
and
(v) PUT {\code /posts/id} : updates the contents of a blog post (body:
the new text of the blog post and the checksum of the older
version of the blog post's text).
To model an imaginary subtle bug,
at every update of a blog post (PUT request with body text and checksum)
the service checks if the checksum provided in the request
matches the recorded checksum for the current blog post, and if it does,
an uncaught exception is raised.
Thus, this bug will be triggered and detected only
if dependencies on dynamic objects shared across requests are taken into account
during test generation. 

\heading{Gitlab.}
We answer (Q2) and (Q3) using \gitlab, an open-source web service for
self-hosted Git. \gitlab's back-end functionality is written in over 376K
lines of ruby code using ruby-on-rails~\cite{rails}.
%
%
% ...
%
It follows the MVC web development paradigm and exposes its functionality over
a REST API. \gitlab's API is extensively documented and has hundreds of
individual API requests spread across \nGitlabAPIGroups groups~\cite{gitlab-doc}.
There is a publicly
available \gitlab Swagger specification which we used for our
experiments~\cite{gitlab-swagger}. A typical GitLab deployment consists of the
following components: (i) a low-level HTTP server used to proxy-pass the Rails
Unicorn web server, (ii) a Sidekiq job queue which, in turn, uses (iii) redis
as a non-persistent database, and (iv) a structured database for persistent
storage. We follow this
deployment and, unless otherwise specified, apply the following
configuration settings: we use Nginx to proxypass the Unicorn web server and
configure \nUnicornWorkers Unicorn workers limited to up to \nUnicornMem of physical memory;
we use postgreSQL for persistent storage configured with a pool of \nPostGREWorkers workers;
we use \gitlab's default configuration for sidekiq queues and reddis workers;
finally, we mount the repositories' destination folder in physical memory.
According to \gitlab's deployment recommendations, our configuration should
scale up to 4,000 concurrent users~\cite{gitlab-requirements}.

\heading{Fuzzing Dictionaries.}
For the experiments in this section, we use the following dictionaries for fuzzable primitives types:
{\em string} has possible values ``sampleString'' and ``'' (empty string);
{\em integer} has possible values ``$0$'' and ``$1$'';
{\em boolean} has possible values ``$true$'' and ``$false$''.

All experiments were run on Ubuntu 16.04 Microsoft Azure VMs configured with eight
Intel(R) Xeon(R) E5-$2673$ v3 @ $2.40$GHz CPU cores and $28$GB of physical
memory, unless otherwise specified.

%% file: sections/04-2-demo-server.tex
\subsection{Techniques for Effective REST API Fuzzing}
\label{sec:eval:benchmarks}
\begin{figure*}[t]
    \begin{minipage}{\textwidth}
    \centering
    \subfigure{
        \includegraphics[width=0.31\textwidth]{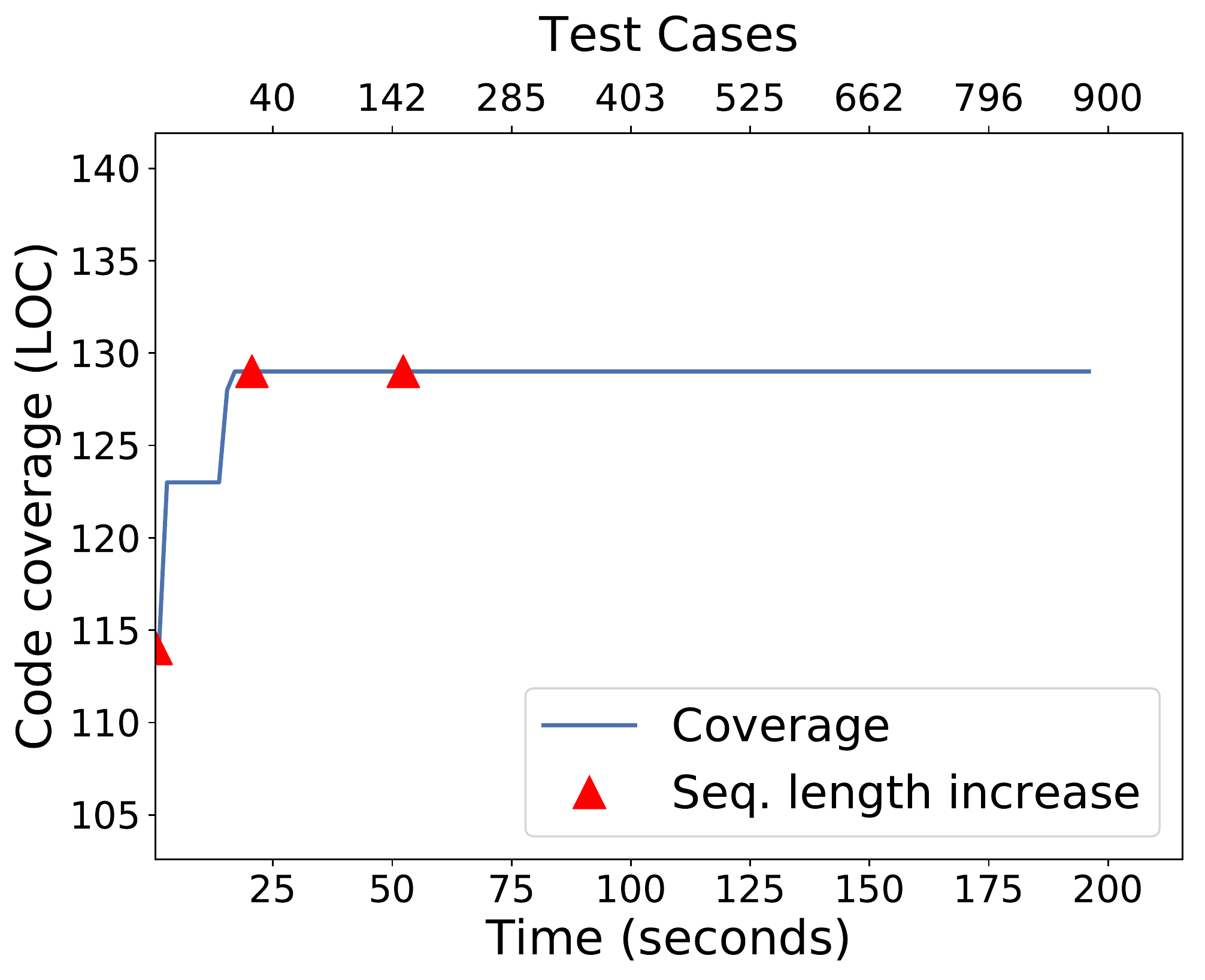}
            }
    \subfigure{
        \includegraphics[width=0.31\textwidth]{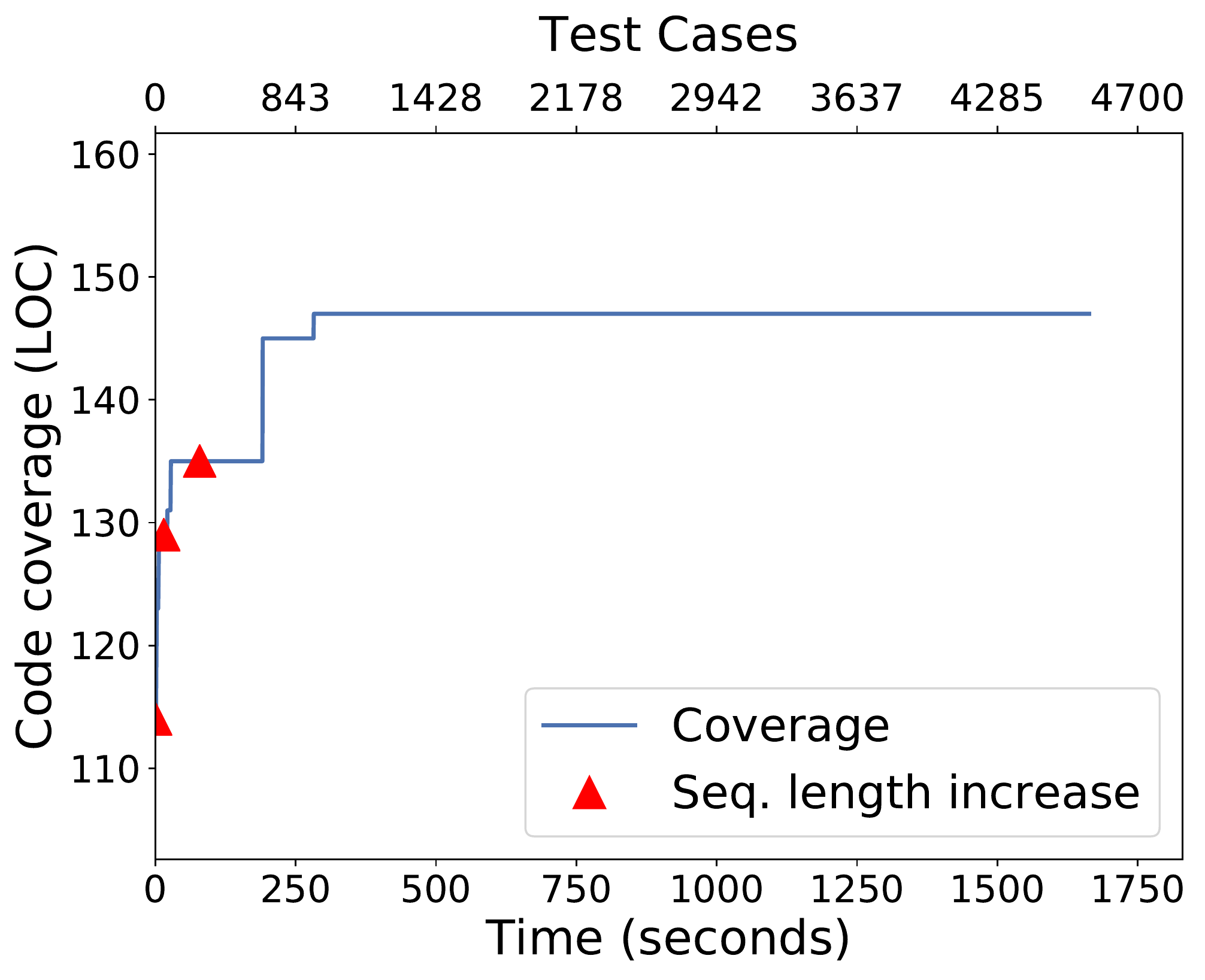}
            }
    \subfigure{
        \includegraphics[width=0.31\textwidth]{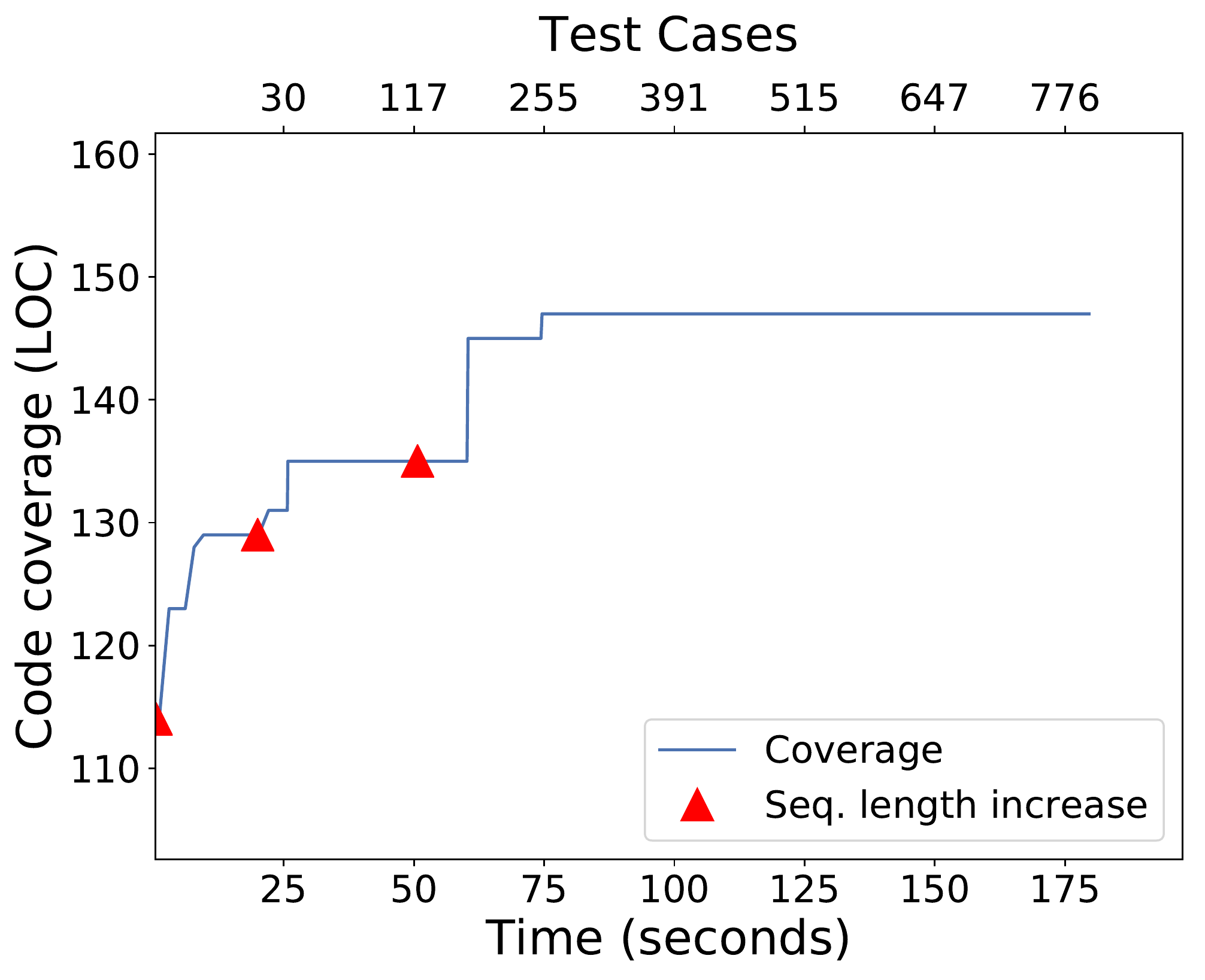}
            }
    \vspace{-8pt}
    \end{minipage}\\[1em]
    \begin{minipage}{\textwidth}
    \centering
    \subfigure{
        \includegraphics[width=0.31\textwidth]{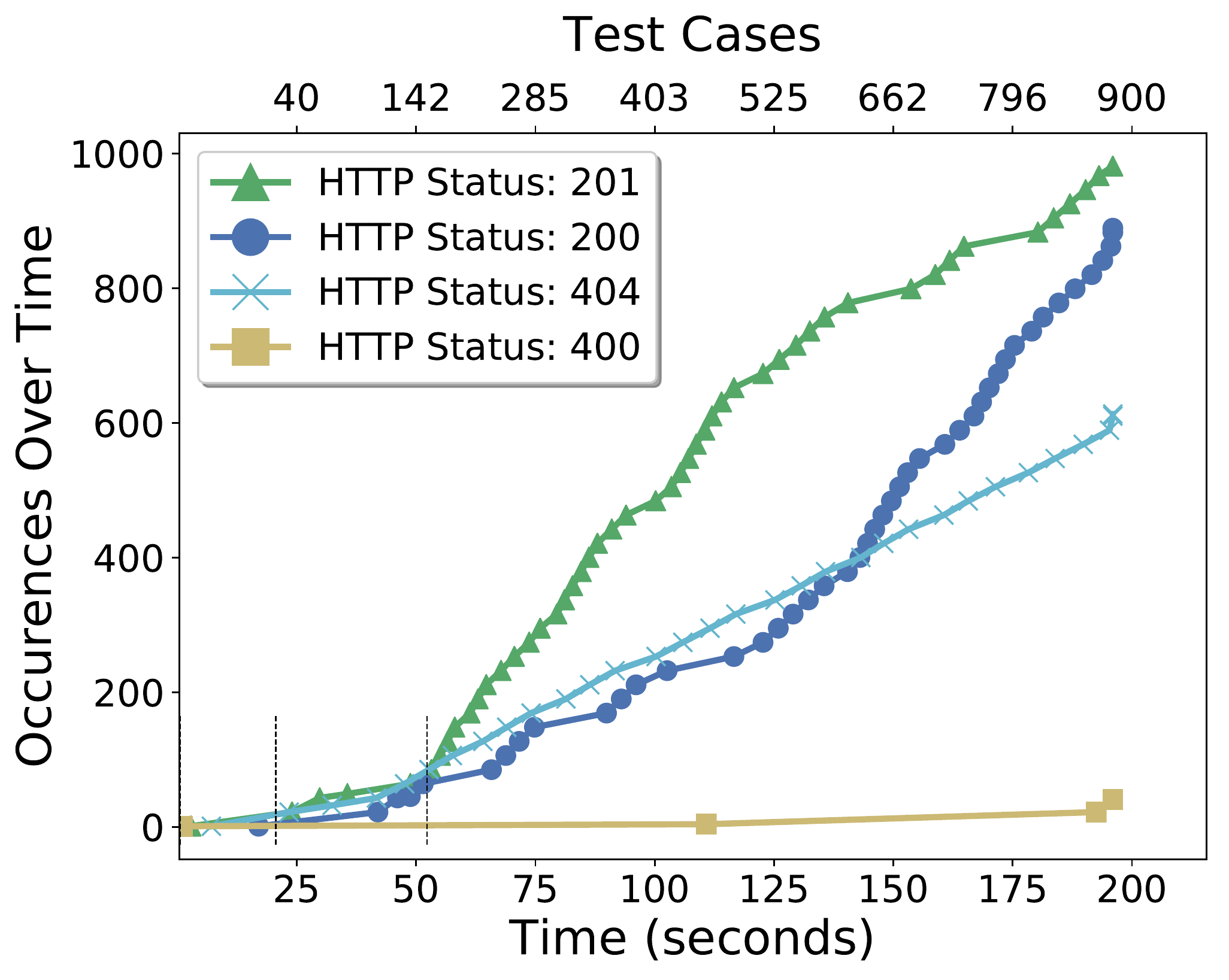}
    }
    \subfigure{
        \includegraphics[width=0.31\textwidth]{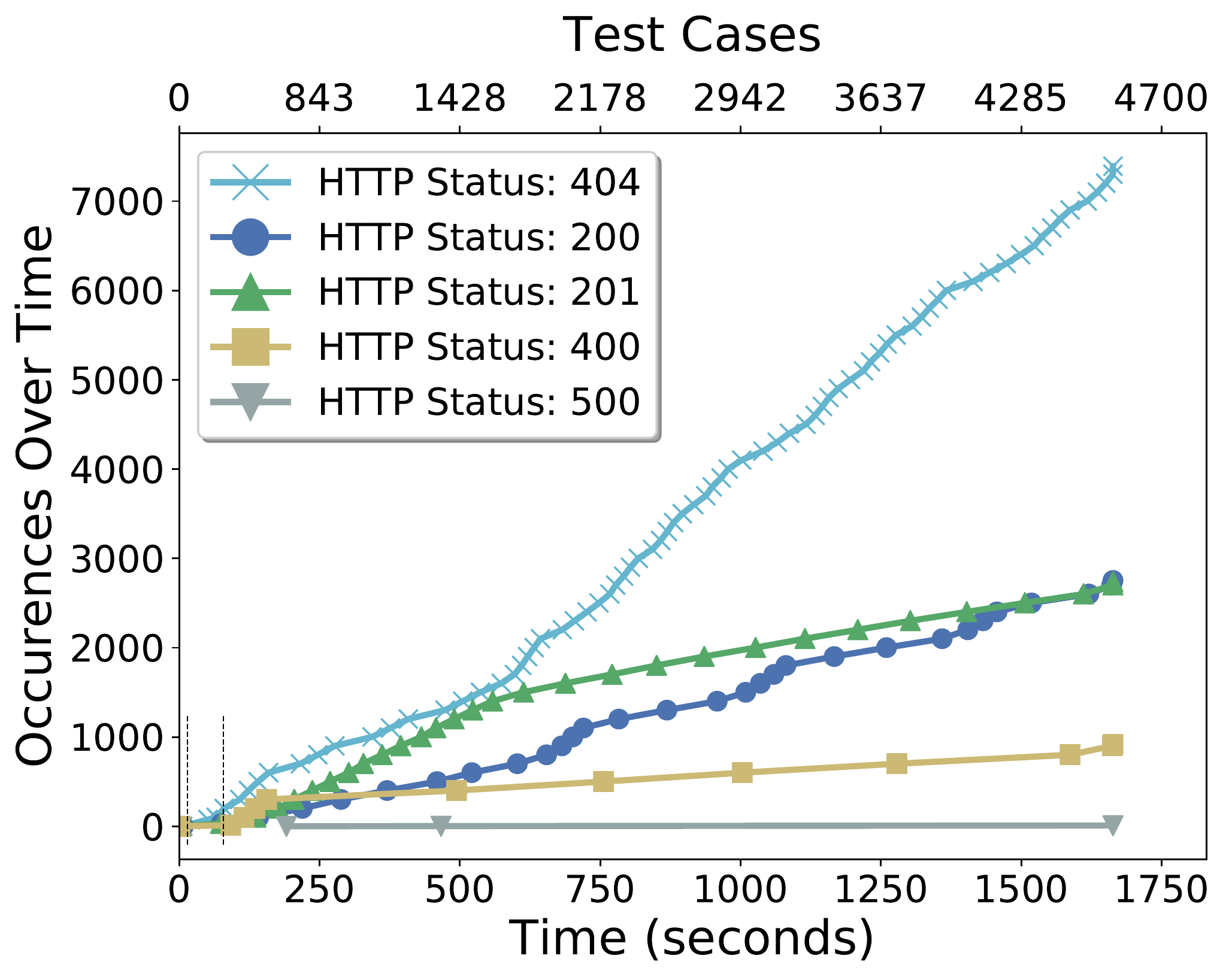}
    }
    \subfigure{
        \includegraphics[width=0.31\textwidth]{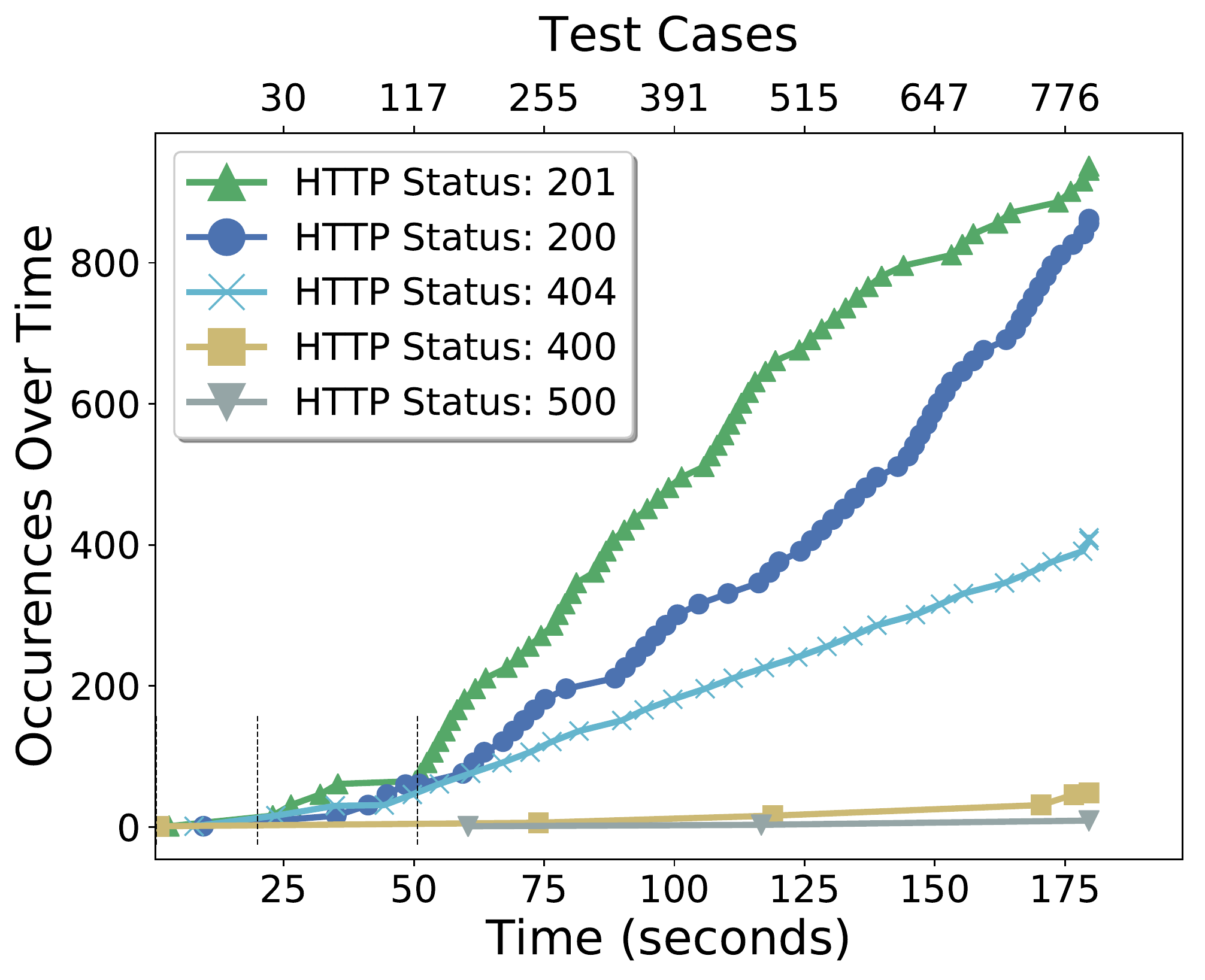}
    }
    \vspace{-10pt}
    \caption{
        {\bf Code Coverage and HTTP Status Codes Over Time.}
        Shows the increase in code coverage over time (top)
        and the cumulative number of HTTP status codes received over time (bottom),
        for the simple blog posts service.
        {\underline {\em Left:}} \restler ignores dependencies among request types.
        {\underline {\em Center:}} \restler ignores dynamic feedback.
        {\underline {\em Right:}} \restler utilizes both dependencies among request types
        and dynamic feedback.  When leveraging both techniques,
        \restler achieves the best code coverage and
        finds the planted {\tt 500} ``Internal Server Error'' bug
        with the least number of tests.
    }
    \label{fig:demoserver}
    \end{minipage}\\[1em]
\end{figure*}

\label{sec:benchmarks}
In this section, we report experimental results with our blog posts
service to determine whether both (1) inferring dependencies among
request types and (2) analyzing dynamic feedback are necessary for
effective automated REST API fuzzing (Q1). We choose a controled
experiment with a known simple blog-posts service in order to clearly
measure and interpret the testing capabilities of the two core
techniques being evaluated. Those capabilities are evaluated by
measuring service code coverage and client-visible HTTP status codes.

Specifically, we compare results obtained when exhaustively generating
all possible request sequences of length up to three, with three
different test-generation algorithms:
\begin{enumerate}
\topsep0pt
\itemsep0pt
\item \restler ignores dependencies among request types and treats dynamic objects  -- such as post
{\code id} and {\code checksum} -- as fuzzable primitive type {\tt
string} objects, while still analyzing dynamic feedback.
\item \restler ignores service-side dynamic feedback and does not eliminate invalid
request sequences during its search, but still infers dependencies
among request types and generates request sequences satisfying such
dependencies.
\item \restler follows the algorithm of Figure~\ref{fig:main-algorithm} and 
uses both dependencies among request types and dynamic feedback.
\end{enumerate}
\F\ref{fig:demoserver} shows the number of tests, \ie request sequences, generated up 
to maximum length 3 by each of these three algorithms, from left to
right. At the top, the figure shows cumulative code coverage measured
in lines of python code (using python's {\code settrace} system
utility) and increasing over time, as well as when the sequence length
increases (from 1 to 3). At the bottom, the figure shows the
cumulative number of HTTP status code received as responses so far.

\heading{Code Coverage.}
First, we observe that without considering dependencies among request
types (\F\ref{fig:demoserver}, top left), code coverage is limited to up to $130$~lines and there
is no increase over time, despite increasing the length of request
sequences. This is expected and illustrates the limitations of a naive
approach to effectively test a service where values of dynamic objects
like {\tt id} and {\tt checksum} cannot be randomly guessed or picked
among values in a small predefined dictionary.  On the other hand,
by infering dependencies among requests manipulating such objects
\restler achieves an increase in code coverage up to $150$~lines of code
(\F\ref{fig:demoserver}, top both center and right).  

Second, we can see that without considering dynamic feedback to prune
invlid request sequences in the search space (\F\ref{fig:demoserver}, top center), the number of
tests generated grows quickly, even for a simple API. Specifically,
without considering dynamic feedback (\F\ref{fig:demoserver}, top center),
\restler produces more that $4,600$ tests that take $1,750$ seconds and cover
about $150$ lines of code; in contrast, by considering dynamic
feedback (\F\ref{fig:demoserver}, top right), the state space is significantly reduced and
\restler can achieve the same code coverage with less than $800$ test
cases and only $179$ seconds.

\heading{HTTP Status Codes.}
%
%
%
%
We make two observations.  First, focusing on {\tt 40X} status codes,
we notice a high number of {\tt 40X} responses when ignoring dynamic
feedback (\F\ref{fig:demoserver}, bottom center).  This is expected
since without considering service-side dynamic feedback, the number of
possible invalid request sequences grows quickly. In contrast,
considering dynamic feedback dramatically decreases the percentage of
{\tt 40X} status codes from $60\%$ to $26\%$ (\F\ref{fig:demoserver},
bottom left) and $20\%$ (\F\ref{fig:demoserver}, bottom right),
respectively, without or with using dependencies among request
types. Moreover, when also using dependencies among request types
(\F\ref{fig:demoserver}, bottom right), we observe the highest
percentage of {\tt 20X} status codes (approximately $80\%$),
indicating that \restler then exercises a larger part of the service
logic, as also confirmed by coverage data (\F\ref{fig:demoserver}, top
right).

Second, when ignoring dependencies among request types, we see that no
{\tt 500} status code are detected (\F\ref{fig:demoserver}, bottom
left), while \restler finds a handful of {\tt 500} status codes when
using dependencies among request types (see (\F\ref{fig:demoserver},
bottom left and bottom right). These {\tt 500} responses are triggered
by the unhandled exception we planted in our blog posts service after
a PUT blog update request with a checksum matching the previous blog
post's body (see Section~\ref{sec:subsec:setup}). As expected,
ignoring dependencies among request types does not find this bug
(\F\ref{fig:demoserver}, bottom left). In contrast, analyzing
dependencies across request types and using the checksum returned by a
previous GET {\code /posts/id} request in a subsequent PUT {\code
/posts/id} update request with the same {\tt id} does trigger the
bug. Morever, when also using dynamic feedback, the search space is
pruned while preserving this bug, which is then found with the least
number of tests (\F\ref{fig:demoserver}, bottom right).

Overall, these experiments illustrate the complementarity between
using dependencies among request types and using dynamic feedback, and
show that both are needed for effective REST API fuzzing.

%% file: sections/04-3-gitlab.tex
\subsection{Deeper Service Exploration}
\label{sec:eval:gitlab}
\begin{table}[t]
{
    \scriptsize
    \def\arraystretch{1.2}
    \begin{center}

        \begin{tabular}{
                p{0.85cm}|p{0.81cm}|p{0.4cm}p{0.95cm}p{0.6cm}
                p{0.6cm}p{.85cm}
        }
            \hline
            {\bf API} &{\bf Requests Tested}
            &{\bf Seq. Len.} &{\bf Coverage Increase} &{\bf Tests}
            &{\bf seqSet Size} &{\bf Dynamic Objects}
            \\
            \hline
            \hline
			\multirow{1}{*}{\bf Commits}
			                & 4 / 10 & 1 & 498 & 1 & 1  & 1\\
			                &        & 2 & 959 & 66 & 5 & 66 \\
			                &        & 3 & 1453 & 483 & 93 & 900 \\
			                &        & 4 & 1486 & 7300 & 2153 & 18663 \\
			\cline{3-5}
			\hline
			\multirow{1}{*}{\bf Branches}
			                & 5 / 7 & 1 & 498 & 1 & 1  & 1\\
			                &       & 2 & 926 & 5 & 3 & 5 \\
			                &       & 3 & 969 & 23 & 14 & 37 \\
			                &       & 4 & 986 & 139 & 82 & 289 \\
			                &       & 5 & 988 & 911 & 516 & 2119 \\
			                &       & 6 & 1026 & 4700 & 5652 & 12371 \\
			\cline{3-5}
			\hline
			\multirow{1}{*}{\bf Issues}
			                & 14 / 25 & 1 & 498 & 1 & 1  & 1\\
			                &         & 2 & 879 & 514 & 257 & 514 \\
			                &         & 3 & 1158 & 8400 & 6964 & 16718 \\
			\cline{3-5}
			\hline
			\multirow{1}{*}{\bf Repos}
			                & 3 / 11 & 1 & 498 & 1 & 1  & 1\\
			                &        & 2 & 871 & 18 & 5 & 18 \\
			                &        & 3 & 937 & 231 & 25 & 640 \\
			                &        & 4 & 954 & 1424 & 125 & 4611 \\
			                &        & 5 & 1030 & 5200 & 909 & 20713 \\
			\cline{3-5}
			\hline
			\multirow{1}{*}{\bf Groups}
			                & 10 / 20 & 1 & 687 & 25 & 25  & 1\\
			                &         & 2 & 718 & 1250 & 1225 & 1226 \\
			                &         & 3 & 798 & 7900 & 13421 & 12466 \\
			\cline{3-5}
			\hline
            \hline
        \end{tabular}
        \caption{{\bf Testing Common \gitlab APIs with \restler.}
            Shows the increase in sequence length, code coverage, tests executed, {\tt
            seqSet} size, and the number of dynamic objects being created, until
            a $5$-hours timeout is reached.
            Longer request sequences gradually increase service-side code coverage.
        }
        \label{tab:gitlab-api}
    \end{center}
}
\end{table}

In this section, we use \gitlab to determine whether tests generated by \restler
exercise more service-side logic and code as
sequence length increases (Q2).
In total, \gitlab has $\nGitlabAPIGroups$ API groups
and \restler identifies $\nGitlabSwaggerRequests$ API request types
after analyzing \gitlab's Swagger
specification~\cite{gitlab-doc,gitlab-swagger}. Here, we
constrain our investigation to five of \gitlab's API groups,
related to usual operations with commits, branches,
issues and issue notes,
repositories and repository files, and groups and group membership.
We target $36$ out of $73$ request types defined for the API groups
under investigation, and also include a POST {\code /projects}
request, which is a common root dependency for all five API groups.
Indeed, we focus on request types with POST, PUT, and DELETE HTTP
methods, which may modify the service's internal state by creating,
updating and deleting resources, but we omit request types with GET
methods which are here simple accessors and cause an unnecessary state
space explosion.

For each API group, \T\ref{tab:gitlab-api} gives the number of request
types included in that API as well as the number of request types
tested by \restler within the API (second column). The Table presents
results of experiments performed with all five \gitlab sub-APIs by the
main test-generation algorithm of Figure~\ref{fig:main-algorithm}
(thus using BFS) with a $5$-hours timeout and while limiting the
number of fuzzable primitive-types combinations to maximum $1,000$
combinations per request. Between experiments, we reboot the entire
\gitlab service to restart from the same initial state. 
For each API, as time goes by, the Table shows the increase (going
down) in the sequence length, code coverage, tests executed, {\tt
seqSet} size, and the number of dynamic objects being created, until
the $5$-hours timeout is reached.

\heading{Code Coverage.}
We collect code coverage data by configuring Ruby's {\code
Class:~TracePoint} hook to trace \gitlab's {\code service/lib} folder.
\T\ref{tab:gitlab-api} shows the cumulative code coverage achieved
after executed all the request sequences generated by \restler for each sequence length,
or until the $5$-hours timeout expires.
The results are incremental on top of \nBootCodeCoverage lines of code
executed during service boot.

\HiddenNote{ commented out since we don't have an appendix anymore
We complement these results with comprehensive graphs showing how
coverage evolves over time in \S\ref{sec:appendix}. Note that our purpose
is not to compare \restler;s code coverage against \gitlab
developers' test cases code coverage. Instead,
we use \gitlab as a means to motivate why \restler test-cases exercise
deeper states of server-side logic, as sequence length increases.
}

From \T\ref{tab:gitlab-api}, we can see a clear pattern across all
five experiments: longer sequence lengths consistently lead to
increased service-side code coverage. This is not surprising,
especially for small sequence lengths, as some of the service
functionality can only be exercised after at least a few requests are
executed.

As an example, consider the \gitlab functionality of ``selecting a
commit''. According to \gitlab's specification, selecting a commit
requires two dynamic objects, a {\em project-id} and a {\em
commit-id}, and the following dependencies of requests is implicit:
(1) a user needs to create a project, (2) use the respective {\em
project-id} to post a new commit, and then (3) select the commit using
its {\em commit-id} and the respective {\em project-id}.  Clearly,
this operation can only be performed by sequences of three requests or
more. For the Commit APIs, note the gradual increase in coverage from
$498$ to $959$ to $1,453$ lines of code for sequence lengths of one,
two, and three, respectively.

As is also expected, for API groups with fewer requests (\eg Commits,
Repos, and Branches), \restler's BFS reaches deeper sequences within
the $5$-hour time budget. Most notably, for the Branches API,
service-side code coverage keeps gradually increasing for sequences of
length up to five, and then reaches $1,026$ lines when the $5$-hours
limit expires. In contrast, for API groups with more requests (\eg
Issues and Groups), \restler generates a higher number of shallow
sequences before reaching the $5$-hours timeout.

\heading{Tests, Sequence Sets, and Dynamic Objects.}
%
%
In addition to code coverage, \T\ref{tab:gitlab-api} also shows the
increase in the number of tests executed, the size of {\tt seqSet}
after the RENDER function returns (in line~10 of
Figure~\ref{fig:main-algorithm}), and the number of dynamic objects
created by \restler in order to get that increased code coverage. 
We can clearly see that all those numbers are quickly growing, because
the search space is rapidly growing as its depth increases and because
of the BFS search strategy used here.

Nevertheless, we emphasize that, without the two key techniques
evaluated in Section~\ref{sec:eval:benchmarks}, this growth would be
much worse. For instance, for the Commit API, the {\code SeqSet} size
is $2,153$ and there are $18,663$ dynamic objects created by \restler
for sequences of length up to four. By comparison, since the Commit
API has four request types with an average of $17$ rendering
combinations, the number of all possible rendered request sequences of
up to length four is already more than $21$ millions, and a naive
brute-force enumeration of those would already be untractable.

Still, even with the two core techniques used in \restler, the search
space explodes quickly, and we evaluate other search strategies next.

\HiddenNote{
Specifically, for the Commits API, with 4 requests and on average 17
(17.6) rendering combinations, there are (4*17)^4 = 21,381,376
possible sequences of length 4.

Note: For the Commits API and length three, if we get seqSet size = 93
after RENDER finishes and we have executed 417 tests, one would expect
most of these tests to be invalid request sequences, which would
explain discarding most of these tests from the seqSet (lines 32-35 of
the algorithm). But if we look at Figure 6, we can see that most (80
of all HTTP status code for the Commits BFS experiment are valid
"201". So why do 417 tests result in a such a small seqSet size of 93?
It is because, in order to reach sequence length 3 it means that for
each test case we get at least two out of three status codes in the
valid class 20X and then we try out the last request. Figure 6 shows
all error codes received when constructing a test case, not only those
triggered by the last request. That's why we have more valid codes here!
}

\begin{table}[t]
{
    \scriptsize
    \def\arraystretch{1.1}
    \begin{center}
       \begin{tabular}{@{\extracolsep{.001pt}}p{0.7cm}p{0.7cm}ccccc}
		\hline
		\multirow{2}{*}{{\bf API}}
            & \multirow{2}{*}{\parbox{0.7cm}{\bf Time (hours)}}
                & \multicolumn{2}{c}{{\bf BFS}}
                & \multicolumn{2}{c}{{\bf RandomWalk}}
                & \multicolumn{1}{c}{{\bf Intersection}}
				\\
				\cmidrule(l{5pt}r{5pt}){3-4}
				\cmidrule(l{8pt}r{8pt}){5-6}
				\cmidrule(l{5pt}r{5pt}){7-7}
			&
                & {\bf Cov.} & {\bf Len.}
                & {\bf Cov.} & {\bf Len.} {\bf (restarts)}
                & {\bf Cov.}
				\\
			\hline
		\hline
		%
		%
		{\bf Commits }    & 1 & 65 & 4 & 0 & 10 & 1420\\
		& 3 & 45 & 4 & 0 & 12 & 1441\\
		& 5 & 44 & 4 & 0 & 14 (116)& 1442\\
		\hline
		{\bf   Branches }    & 1 & 0 & 6 & 0 & 21 & 988\\
		& 3 & 0 & 6 & 0 & 21 & 988\\
		& 5 & 38 & 6 & 0 & 24 (462) & 988\\
		\hline
		{\bf Issues }    & 1 & 0 & 3 & 80 & 9 & 1020\\
		& 3 & 36 & 3 & 3 & 15 & 1119\\
		& 5 & 39 & 3 & 3 & 15 (60)& 1119\\
		\hline
		{\bf Repos }    & 1 & 90 & 5 & 0 & 12 & 940\\
		& 3 & 90 & 5 & 0 & 14 & 940\\
		& 5 & 90 & 5 & 0 & 14 (138)& 940\\
		\hline
		{\bf Groups }    & 1 & 0 & 3 & 31 & 23 & 754\\
		& 3 & 0 & 3 & 25 & 31 & 760\\
		& 5 & 28 & 3 & 16 & 31 (197)& 770\\
		\hline
		\hline
        \end{tabular}
        \caption{
            {\bf Search Strategies and Code Coverage.}
            Compares test results for \gitlab APIs using the
            {\em BFS} and {\em RandomWalk} search strategies, after 1, 3, and 5 hours.
            The table shows the maximum length of request sequences and the unique
	    lines of code covered by each search strategy, as well as their intersection
            (excluding service-boot coverage). For {\em RandomWalk}, the total number
			of restarts after $5$ hours is also shown in parenthesis.
            Although both strategies mostly cover overlapping lines of
            code, some are covered only by one or the other. Overall, after 5 hours, 
            {\em BFS} often generates the best coverage.
        }
        \label{tab:gitlab-api-exploration-methods}
    \end{center}
}
\end{table}
%
%
%
%

%% file: sections/04-4-fuzzing-schemes.tex
\subsection{Search Strategies}
\label{sec:eval:fuzzing-schemes}

We now present results of experiments with the search strategies BFS, RandomWalk, and
BFS-Fast defined in
Section~\ref{sec:restler}. We start by comparing BFS with RandomWalk.

\heading{Code Coverage.}
\T\ref{tab:gitlab-api-exploration-methods} shows the unique lines of code covered by
BFS and RandomWalk, as well as their intersection (excluding
service-boot coverage), after $1$, $3$, and $5$ hours of search.  The
Table also shows the maximum length of request sequences generated so
far and, for RandomWalk, the total number of restarts after $5$
hours in parenthesis.

We observe that both search strategies quickly converge to an
overlapping set of lines of code. After 5 hours of search,
BFS often generates the best coverage. However, in the Issues and
Groups APIs, there are still some lines of code covered only by BFS or
by RandomWalk.

By construction, BFS provides full-grammar coverage whenever it
increases its sequence length, and this feature seems to pay off when
measuring service-side coverage. In contrast, RandomWalk goes deeper
faster but without providing full-grammar coverage. The effects on
coverage of these two different search strategies is more visible on
broader APIs and search spaces with more request types, like the
Issues and Groups APIs (see Table~\ref{tab:gitlab-api}). After $1$
hour of search for these two APIs, RandomWalk has a head-start of,
respectively, $80$ and $31$ lines of code in coverage compared to
BFS. However, after $5$ hours of search, BFS coverage has caught up
with RandomWalk and is now overall better.

\begin{figure}[t]
    \centering
    \includegraphics[width=0.48\textwidth]{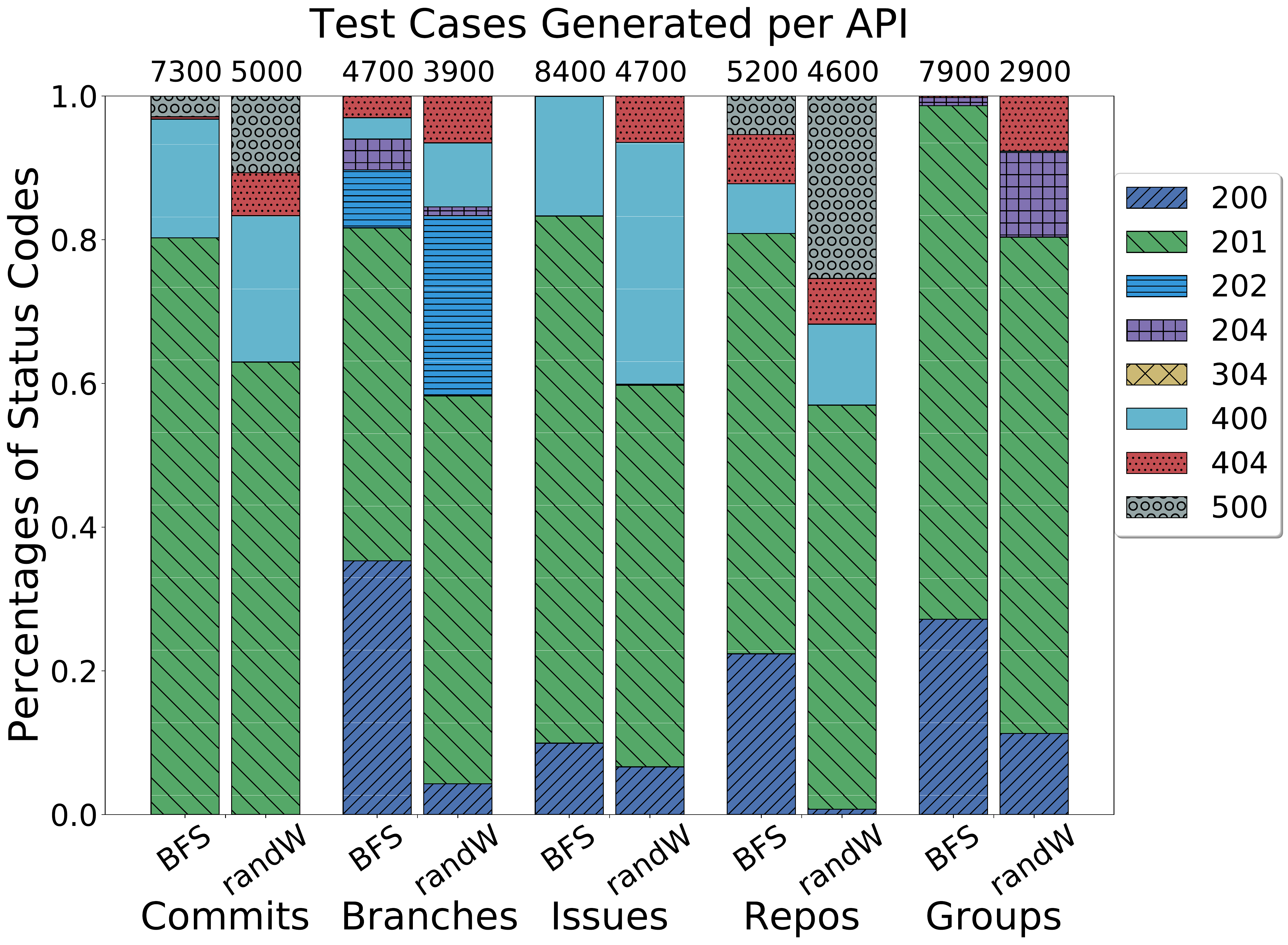}
    \vspace{-12pt}
    \caption{
        {\bf Distribution of HTTP Status Codes for \gitlab APIs.}
        Shows the HTTP status codes collected during testing
        with {\em BFS} and {\em RandomWalk}.
        Different search strategies exercise the service under test differently.
    }
    \label{fig:gitlab_hist}
\end{figure}

\heading{HTTP Status Codes.}
\F\ref{fig:gitlab_hist} shows the distribution of HTTP status codes
collected during $5$ hours of testing of each of the \gitlab APIs with
BFS and RandomWalk. Focusing on each API group separately, we can see
that the distributions obtained with BFS and RandomWalk are
different. Both search strategies seem to exercise the service-under
test differently. For instance, in the Commits, Issues and Groups
APIs, only RandomWalk triggers {\tt 404} status codes. Yet in the
Commits case, BFS's coverage is strictly better (see
Table~\ref{tab:gitlab-api-exploration-methods}). Another interesting
difference is that, in the Commits and Repos APIs, RandomWalk triggers
at least three times more {\tt 500} ``Internal Server Error'' status
codes than BFS. We use those {\tt 500} status codes to detect bugs,
and we will discuss those more in Sections~\ref{sec:eval:usability}
and~\ref{sec:case-studies}.

Overall, the differences in coverage and HTTP status-code distribution
profiles obtained with BFS and RandomWalk motivate us to explore
trade-offs between these two extreme search strategies. Specifically,
we now evaluate such a trade-off, namely BFS-Fast.

\begin{table}[t]
{
    \scriptsize
    \def\arraystretch{1.2}
    \begin{center}
       \begin{tabular}{@{\extracolsep{.02pt}}p{1cm}p{0.7cm}cccccc}
		\hline
		\multirow{2}{*}{{\bf API}}
            & \multirow{2}{*}{\parbox{0.8cm}{\bf Time (hours)}}
                & \multicolumn{3}{c}{{\bf BFS-Fast}}
                & \multicolumn{3}{c}{{\bf BFS}}
				\\
				\cmidrule(l{5pt}r{5pt}){3-5}
				\cmidrule(l{8pt}r{8pt}){6-8}
			&
                & {\bf Len.} & {\bf Cov.} & {\bf seqSet}
                & {\bf Len.} & {\bf Cov.} & {\bf seqSet}
				\\
			\hline
		\hline
		%
		%
		{\bf Commits }      & 1 & 10 & 1482 & - & 4 & 1485 & -\\
		                    & 3 & 14 & 1482 & - & 4 & 1486 & -\\
		                    & 5 & 17 & 1482 & 6 & 4 & 1486 & 2153\\
		\hline
		{\bf   Branches }   & 1 & 22 & 974 & - & 6 & 988 & -\\
		                    & 3 & 42 & 974 & - & 6 & 988 & -\\
                            & 5 & 42 & 974 & 6 & 6 & 1026 &5652\\
		\hline
		{\bf Issues }       & 1 & 4 & 1100 & - & 3 & 1020 & -\\
		                    & 3 & 5 & 1150 & - & 3 & 1155 & -\\
                            & 5 & 5 & 1150 & 272 & 3 & 1158 & 6964\\
		\hline
		{\bf Repos }        & 1 & 12 & 935 & - & 5 & 1030 & -\\
		                    & 3 & 18 & 972 & - & 5 & 1030 & -\\
		                    & 5 & 23 & 986 & 13 & 5 & 1030 & 909\\
		\hline
		{\bf Groups }       & 1 & 13 & 772 & - & 3 & 754 & -\\
		                    & 3 & 19 & 772 & - & 3 & 760 & -\\
		                    & 5 & 21 & 772 & 31 & 3 & 798 & 13421\\
		\hline
		\hline
        \end{tabular}
        \caption{
            {\bf Comparison of {\em BFS-Fast} and {\em BFS} over Time}.
            Shows the maximum sequence length and the
            increase in lines of code covered (excluding service-boot coverage) obtained with each search strategy
             after $1$, $3$, and $5$ hours.
            The {\code seqSet} size is also shown after $5$ hours.
            Although {\em BFS} covers slightly more lines of code, {\em BFS-Fast}
            reaches deeper request sequences and maintains a much smaller {\code seqSet} size.
        }
        \label{tab:bfs-fast}
    \end{center}
}
\end{table}

\heading{Comparison with BFS-Fast.}
Like RandomWalk, BFS-Fast goes deeper faster than BFS. But like BFS,
BFS-Fast still provides full-grammar coverage when increasing sequence
length.

\T\ref{tab:bfs-fast} presents results of experiments comparing BFS-Fast with BFS.
The Table shows the maximum sequence length and the increase in lines
of code covered (excluding service-boot coverage) by each search
strategy after $1$, $3$, and $5$ hours.  The {\code seqSet} size is
also shown after $5$ hours.

We can see that, although BFS covers slightly more lines of code,
BFS-Fast reaches deeper request sequences and maintains a much smaller
{\code seqSet} size. With BFS-Fast, the set {\code seqSet} remains
consistently small: at each iteration when the sequence length
increases, the function EXTEND only creates one new sequence per
request type; then this set is in turn expanded by the function
RENDER, but then shrinks again as many fuzzing combinations lead to
invalid responses. This explains why {\code seqSet} tends to oscillate
and stabilize around small sizes as the search goes deeper.

In practice, controling the size of {\code seqSet} is key to {\em
scalability} when reaching greater depths during longer searches, or
when facing broader search spaces due to larger APIs with more request
types. This is why we adopt BFS-Fast as the default search strategy
used in \restler.

Although code coverage with BFS-Fast is slightly less than with BFS
after $5$ hours of search, BFS-Fast actually detects all the {\tt 500}
HTTP status codes found by BFS within $5$ hours, as well as those
found by RandomWalk, as discussed in the next section.

%% file: sections/04-5-usability.tex
\subsection{Bug Bucketization}
\label{sec:eval:usability}

Before discussing real errors found with \restler, we introduce a
bucketization scheme to cluster the numerous {\tt 500} ``Internal
Server Errors'' that can sometimes be reported. Indeed, as usual when
fuzzing, different instances of a same bug can be repeatedly found
over and over again. Since all the bugs found have to be inspected by
the user, it is therefore important in practice to facilitate this
analysis by identifying likely-redundant instances of a same unique
bug.

In our context, we define a {\em bug} as a {\tt 500} HTTP status code being
received after executing a request sequence. Thus, every bug found is
associated with the request sequence that was executed to find
it. Given this property, we use the following bucketization procedure for
the bugs found by \restler:
\begin{quote}
Whenever a new bug is found during the search, we compute all
non-empty suffixes of its request sequence (starting with the smallest
one), and we check whether some suffix is a previously-recorded
request sequence leading to a bug found earlier during the search. If
there is a match, the new bug is added to the bucket of that previous
bug. Otherwise, a new bucket is created with the new bug and its
request sequence.
\end{quote}
Note that, in the above procedure, requests in request sequences are
identified by their type, not by how they are rendered -- fuzzable
pritimitive types are not taken into account, and requests rendered
differently are always considered equivalent. For a request sequence
of length $n$, there are $n$ suffixes. When using BFS or BFS-Fast,
this bucketization scheme will identify bugs by the shortest request
sequence needed to find it.

For the $5$-hours experiments with GitLab reported earlier in this section,
\restler found bugs ({\tt 500} HTTP status codes) in two out of the five \gitlab API groups
(see Figure~\ref{fig:gitlab_hist}). After bucketization, there are two
buckets found for the Commits API, two buckets found for the Repos
API, and none for the Branches, Issues, and Groups APIs. Note that all
four buckets are found within $5$ hours by BFS, RandomWalk, and
BFS-Fast.

\newbox\restlerBucketSample
\begin{lrbox}{\restlerBucketSample}
\begin{lstlisting}[linewidth=215pt,language=bucketlog]
1/2: POST /api/v4/projects HTTP/1.1
Accept: application/json
Content-Type: application/json
Host: 127.0.0.1
PRIVATE-TOKEN: [FILTERED]
'{"name":restler_fuzzable("string")}'

2/2: POST /api/v4/projects/projectid/repository/commits HTTP/1.1
Accept: application/json
Content-Type: application/json
Host: 127.0.0.1
PRIVATE-TOKEN: [FILTERED]
'{
   "branch": restler_fuzzable("string"),
   "commit_message": restler_fuzzable("string"),
   "actions":
   [
     {
       "action": ["create", "delete", "move", "update"],
       "file_path": restler_fuzzable("string"),
       "content":  restler_fuzzable("string"),
    }
  ]
}'
\end{lstlisting}
\end{lrbox}

\begin{figure}[t]
    \center
        \usebox\restlerBucketSample
    \caption{
        {\bf Sample \restler Bug Bucket.} This request sequence identifies
        a unique bug found in \gitlab's Commits API. The sequence
        consists of two requests: the first POST request creates a
        project, and the second POST request uses the id of that new project and posts a
        commit with some action and some file path.
    }
    \label{fig:testcase_bucket}
\end{figure}

\F\ref{fig:testcase_bucket} depicts the (unrendered) request sequence of length $2$ identifying
one of the bug buckets found for the Commits API. This request
sequence triggers a {\tt 500} ``Internal Server Error'' when a user creates
a project and then posts a commit with an empty branch name -- that is,
when the second request is rendered with its branch field being
the empty string.  This specific error can be reached by numerous
longer sequences (\eg creating two projects and posting a commit with
an empty branch name) and multiple value renderings (\eg any feasible
``content'' rendering along with an empty branch name). Nevertheless,
the above bucketization procedure will group together all those
variants into the same bucket.

We discuss the details causing this error as well as other unique bugs
in the following section. Indeed, \restler also found unique bugs in
the Groups, Branches, and Issues GitLab APIs when running longer
fuzzing experiments.

\HiddenNote{KEEP THIS HERE? OR USE IT AS INTRO TO NEXT SECTION INSTEAD?
Overall, our evaluation demonstrated that \restler intelligently
prunes the search space and constructs test-cases that uncover logical
errors within reasonable time-frame.  In what follows, we present case
studies of new, unknown logical errors found with \restler.}

%% file: sections/05-case-studies.tex
\section{New Bugs Found in GitLab}
\label{sec:case-studies}

In this section we discuss ``unique bugs'' (\ie after bucketization)
found so far by \restler during experiments with the GitLab APIs. Note
that \restler does not report false alarms and that all unique bugs
were so far rather easily reproducible, unless otherwise specified.
All the bugs reported in this section were found by \restler while running
for at most 24 hours.

Early in our experiments, \restler found a first bug in the Branches
API when using a fuzzing dictionary which included a (1-byte)
``$\backslash 0$'' string as one of its values.
The bug seems due to a parsing issue in
{\code ruby-grape}, a {\code ruby} middleware library
for creating REST APIs~\cite{grape}.
Since other \gitlab APIs also depend on the {\code ruby-grape} library,
the bug is easy to trigger when using a dictionary with a ``$\backslash 0$''
string.
In other words, many bugs in
different buckets as defined in the previous section point to that
same common root cause. In order to eliminate this
``noise'' in our experiments, we subsequently removed that dictionary
value for the experiments of Section~\ref{sec:evaluation} and for those
that led to discover the bugs discussed below. We now describe additional
bugs found per API group.

\heading{Commits.}
In the Commits API, \restler found three bugs. The first one is
triggered by the request sequence shown in
Figure~\ref{fig:testcase_bucket} when the branch name in the second
request is set to an empty string.  According to \gitlab's
documentation~\cite{gitlab-doc}, users can post commits with multiple
files and actions, including action ``create``, which creates a file
with a given content on a selected target branch.
For performance benefits, \gitlab uses {\code rugged} bindings
to native {\code libgit2}, a pure C implementation of the Git core
methods~\cite{libgit}.  Due to incomplete input validation, an invalid
branch name, like an empty string, can be passed between the two
different layers of abstraction as follows.
The ruby code checks if the target branch exists  by invoking
a native C function whose return
value is expected to be either NULL or an existing entry. However, if
an unmatched entry type (\eg an empty string) is passed to the C function,
an exception is
raised. Yet, this exception is unexpected/unhandled by the
higher-level ruby code, and thus causes a {\tt 500} ``Internal Server
Error''. The bug can easily be reproduced by creating a project and
posting a commit with action ``create'' to a 
branch whose name is set to the empty string.

%
%
%

A second bug found by \restler seems similar: (1) create a project,
(2) post a valid commit with action ``create'', and then (3) cherry-pick
the commit to a branch whose name is set to the empty string.  Both
bugs seem due to similar improper validation of branch names.

A third bug found by \restler is slightly different: (1) create a
project, and then (2) create a commit with action ``move'' and omit the
``previous\_path'' field from the action's parameters.
In this case, the bug is due to an incomplete
validation of the ``previous\_path'' field. We do not know how severe this
specific bug is, but there have been past security vulnerabilities in
GitLab due to improper file-paths validation, which could have been
exploited by an attacker to perform unauthorized move operations and
leak private data~\cite{gitlabcves}.

\heading{Repos.}
In the Repos API, \restler found two
bugs. The first one is triggered when a user attempts to
(1) create a project, and then (2) create a file using a dictionary of parameters
in which either the author's email or the author's name (optional parameters)
is set to the empty string. The bug seems due to incomplete input validation
and error propagation between different layers of abstraction. Specifically,
ruby code invokes native code from {\code libgit2} which attempts
to parse commit options and create a commit signature,
using the author's name and email. In case an
empty string is encountered in the author's name of email field,
an unexpected/unhandled exception is raised to the ruby code causing
a {\tt 500} ``Internal Server Error''.

The second bug is similar, but requires a deeper sequence of actions in order
to be triggered:
(1) create a project, (2) create a file, and (3) edit the file content with
an empty author email or author name. Note that both bugs 
seem to be due to the same piece of code.

\heading{Groups.}
For the Groups API, \restler found a bug which can be triggered
by creating a new group using an invalid parent group id.
According to \gitlab's documentation, users can create new groups and optionally
define group nesting by providing a parent group id through a 
parameter named ``parent\_id''. If group nesting is defined,
the ACLs of the new group must be at least as restricted as those
of the parent. However, the ruby code responsible for enforcing such a
relationship between ACLs uses ``parent\_id'' to access
the respective ACLs without checking whether ``parent\_id''
lies within a range of valid integers for {\code activerecord}.
Therefore, using a large integer value in the field of the ``parent\_id'' param
causes a {\code RangeError} exception.
Furthermore, after manual inspection, we realized that assigning to
``parent\_id'' any integer value which is currently not assigned to an
existing group will also cause a {\code NilClass} excetion. This is because
there is always an attempt to
access the ACLs of an object without first checking that the target object
actually exists.

\heading{Branches and Issues.}
Finally, for both the Branches and the Issues APIs
\restler found {\tt 500} ``Internal Server Errors'' when running in
parallel fuzzing mode, where multiple request sequences are
being tested concurrently. For instance, one such
bug can be reproduced as follows:
(1) create a project, (2) create a valid branch and
{\em at the same time} (\eg using a second process)
create another branch on the same project.
To successfully reproduce this bug, one needs to make the two requests
simultaneously in order for both of them to be within a small time window
in which some internal book-keeping is being performed.
These bugs seem due to concurrency issues and may be related to
the attacks on database-backed web applications
presented by Warszawski et al.~\cite{ACIDRain}.

\heading{Discussion.}
From the description of the bugs found so far with \restler we see an emerging
two-fold pattern. First, \restler produces a sequence of requests that
exercise a service deep enough so that it reaches a particular interesting and valid ``state''.
Second, while the service under test is in such a state,
\restler produces an additional request with an unexpected fuzzed primitive
type -- like a string containing ``$\backslash 0$'' or an empty string.
Therefore, triggering a bug requires a combination of these two features.
We believe that \restler already does a good job at exercising a service deep
enough, thanks to its dependency analysis. However,
more fuzzing values and more properties to
check in responses could boost \restler's bug-finding capabilities -- 
there is clearly room for improvement here.

We emphasize that the findings reported in this section are still
preliminary. The bugs found so far are input validation bugs whose
severity could range from mere crashes (which might only cause
Denial-Of-Service attacks in the worst case) to more subtle
access-control bugs (which might cause service state corruption or
private-information disclosure). At the time of this writing, we are
in communication with the GitLab developers, and we will update the
next version of this paper with their feedback.

%% file: sections/06-related-work.tex
\section{Related Work}
\label{sec:relwork}

\HiddenNote{
Should \restler be called a fuzzer? In contrast to other fuzzing
tools, \restler does not require seed inputs which are then
mutated/fuzzed. Also, how security-critical are the bugs \restler can
find? Are they general input-validation or back-end logic bugs? How
severe are they? [ we need to discuss this ]

[ Advantages of calling \restler a fuzzing tool: (1) this would
indicate that REST-ler can find security-related bugs and (2) people
would therefore pay more attention to it (and would perhaps one day be
required to use it, for instance for security compliance
reasons). However, before we can do it, we need evidence REST-ler is
security relevant. Let's use this as a challenge for us over the next
few weeks: can we upgrade REST-ler from a testing tool to a fuzzing
tool by demonstrating that (a) it is effective in finding bugs in
GitLab and (b) that those bugs are security relevant? ]

We need to finish the section on 'new bugs found' and notify the
GitLab owners but, based on the resutls so far, I think we have enough
evidence to call \restler a fuzzer based on the 2 criteria above. 
}

The lightweight static analysis of Swagger specifications done by
\restler (see Section~\ref{sec:overview}) in order to infer dependencies 
among request types is similar to the analysis of type dependencies
performed by the Randoop algorithm~\cite{randoop} for typed
object-oriented programs. However, unlike in the Randoop work, dynamic
objects in Swagger specifications are untyped, and \restler has to
infer those somehow, as best as it can. Sometimes, user help may be
required to unblock \restler when a Swagger specification is not
complete, for instance, when a type of resource is not described in
the Swagger specification itself, such as authentication and
authorization tokens for allowing a test client to access a
service. In the future, it would be interesting to allow users to
annotate Swagger specifications and to declare service-specific types
as well as their properties, in the spirit of code
contracts~\cite{Mey92,BFL10}.

The dynamic feedback \restler uses to prune invalid responses from the
search space (see line~32 in Figure~\ref{fig:main-algorithm}) is also
similar to the feedback used in Randoop~\cite{randoop}. However, the
Randoop search strategy (in particular, search pruning and ordering)
is different from the three simple strategies considered in our work,
namely BFS, BFS-Fast and RandomWalk (the latter being the closest to
Randoop). Moreover, some of the optimizations of the Randoop search
algorithm (related to object equality and filtering) are not directly
applicable and relevant in our context. Of course, other search
strategies could be used, and it would be worth exploring those in our
context in future work.

Our BFS-Fast search strategy is inspired by test generation algorithms
used in model-based testing~\cite{utting2012}, whose goal is to
generate a minimum number of tests covering, say, every state and
transition of a finite-state machine model (\eg see~\cite{YL91}) in
order to generate a test suite to check conformance of a (blackbox)
implementation with respect to that model. It is also related to
algorithms for generating tests from an input grammar while covering
all its production rules~\cite{lammel2006ccc}. Indeed, in our context,
BFS-Fast provides, by construction, a full grammar coverage up to the
given current sequence length. The number of request sequences it
generates is not necessarily minimal, but that number was always
small, hence manageable, in our experiments so far.

Since REST API requests and responses are transmitted over HTTP,
HTTP-fuzzers can be used to fuzz REST APIs. Such fuzzers, like
Burp~\cite{burp}, Sulley~\cite{sulley}, BooFuzz~\cite{boofuzz}, or the
commercial AppSpider~\cite{appspider} or Qualys's
WAS~\cite{qualysWAS}, can capture/replay HTTP traffic, parse HTTP
requests/responses and their contents (like embedded JSON data), and
then fuzz those, using either pre-defined
heuristics~\cite{appspider,qualysWAS} or user-defined
rules~\cite{sulley,boofuzz}. Tools to capture, parse, fuzz, and replay
HTTP traffic have recently been extended to leverage Swagger
specifications in order to parse HTTP requests and guide their
fuzzing~\cite{appspider,qualysWAS,tnt-fuzzer,apifuzzer}. Compared to
those tools, the main originality of \restler is its global dependency
analysis of Swagger specifications and its ability to intelligently
generate sequences of requests without pre-recorded HTTP traffic.

General-purpose (\ie non-Swagger specific) grammar-based fuzzers,
like Peach~\cite{Peach} and SPIKE~\cite{SPIKE}, among
others~\cite{fuzzing-book}, can also be used to fuzz REST APIs. With
these tools, the user directly specifies an input grammar, often
encoded directly by code specifying what and how to fuzz, similar to
the code shown on the right of
Figure~\ref{fig:demoserver_swagger_restler_snippets}. Compared to
those tools, \restler generates automatically an input grammar from a
Swagger specification, and its fuzzing rules are determined separatety
and automatically by the algorithm of Figure~\ref{fig:main-algorithm}.

How to learn automatically input grammars from input samples is
another complementary research
area~\cite{autogram,bastani,GHS17}. \restler currently relies on a
Swagger specification to represent a service's input space, and it
learns automatically how to prune invalid request sequences by
analyzing service responses at specific states. Still, a Swagger
specification could be further refined given representative (unit
tests) or live traffic in order to focus the search towards specific
areas of the input space. For services with REST APIs but no Swagger
specification, it would be worth investigating how to infer it
automatically from runtime traffic logs using machine learning, or by
a static analysis of the code implementing the API.

Grammar-based fuzzing can also be combined~\cite{MX07,GKL08} with
whitebox fuzzing~\cite{SAGE}, which uses dynamic symbolic execution,
constraint generation and solving in order to generate new tests
exercising new code paths. In contrast, \restler is currently purely
blackbox: the inner workings of the service under test are invisible to
\restler which only sees REST API requests and responses. Since cloud
services are usually complex distributed systems whose components are
written in different languages, general symbolic-execution-based
approaches seem problematic, but it would be worth exploring this
option further. For instance, in the short term,
\restler could be extended to take into account alerts (\eg
assertion violations) reported in back-end logs in order to increase
chances of finding interesting bugs and correlating them to specific
request sequences.

The GitLab concurrency-related bugs we reported in the previous
section look similar to known classes of bugs and attacks in
database-backed applications~\cite{ACIDRain}. These bugs are due to
concurrent database operations that are not properly encapsulated in
serializable transactions. When a REST API exposes concurrent database
operations without properly isolating those, data corruptions and
leaks may happen. It would be interesting to develop
concurrency-specific fuzzing rules inspired by the attack techniques
of~\cite{ACIDRain} and to experiment with those in \restler.

In practice, the main technique used today to ensure the security of
cloud services is so-called ``penetration testing'', or {\em pen
testing} for short, which means security experts review the
architecture, design and code of cloud services from a security
perspective. Since pen testing is labor intensive,
it is expensive and limited in scope and depth. Fuzzing tools like
\restler can partly automate and improve the discovery of specific 
classes of security vulnerabilities, and are complementary
to pen testing.

%% file: sections/07-conclusion.tex
\section{Conclusions}
\label{sec:conclusion}

We introduced \restler, the first automatic intelligent tool for
fuzzing cloud services through their REST APIs. \restler analyzes a
Swagger specification of a REST API, and generates tests intelligently
by inferring dependencies among request types and by learning invalid
request combinations from the service's responses. We presented
empirical evidence showing that these techniques are necessary to thoroughly
exercise a service while pruning its large search space of possible
request sequences. We also evaluated three different search strategies
on GitLab, a large popular open-source self-hosted Git
service. Although \restler is still an early prototype, it was already
able to find several new bugs in GitLab, including
security-related ones.

While still preliminary, our results are encouraging. How general are
these results? To find out, we need to fuzz more services through
their REST APIs, add more fuzzing rules to further confuse and trip
service APIs, and check more properties to detect different kinds of
bugs and security vulnerabilities. Indeed, unlike buffer overflows in
binary-format parsers, or use-after-free bugs in web browsers, or
cross-site-scripting attacks in web-pages, it is still largely unclear
what security vulnerabilities might hide behind REST APIs. While past
human-intensive pen testing efforts targeting cloud services provide
evidence that such vulnerabilities do exist, this evidence is still
too anecdotical, and new automated tools, like \restler, are needed
for more systematic answers. How many bugs can be found by fuzzing
REST APIs? How security-critical will they be? This paper provides a
clear path forward to answer these questions.

\HiddenNote{ PG: I THINK THE FOLLOWING IS OBVIOUS - LET'S STAY CRISP AND DON'T MENTION IT...
We could go this with other API
specifications than Swagger, but the approach developed in this paper
for automatically fuzzing cloud services through their APIs relies
fundamentally on leveraging some sort of global specification of that
API.
}